\documentclass{aastex631}

\newcommand{\CII}{[C\,{\sc ii}] }
\newcommand{\OI}{[O\,{\sc i}] }

\begin{document}

\title{Ring or no ring -- Revisiting the Multiphase Nuclear Environment in M31}

\author[0000-0002-7172-6306]{Zongnan Li}
\email{zongnan.li@astro.nao.ac.jp}
\affiliation{National Astronomical Observatories, Chinese Academy of Sciences, A20 Datun Road, Chaoyang District, Beijing, 100101, China}
\affiliation{National Astronomical Observatory of Japan, 2-21-1 Osawa, Mitaka, Tokyo, 181-8588, Japan}
\affiliation{East Asian Core Observatories Association (EACOA) Fellow}
\author[0000-0002-6738-3259]{Zhao Su}
\author[0000-0001-8586-1253]{Sumin Wang}
\author[0009-0006-0194-6211]{Yufan F. Zhou}
\author[0000-0003-0355-6437]{Zhiyuan Li}
\email{lizy@nju.edu.cn}
\affiliation{School of Astronomy and Space Science, Nanjing University, Nanjing 210023, China}
\affiliation{Key Laboratory of Modern Astronomy and Astrophysics, Nanjing University, Nanjing 210023, China}
\author{Xuanyi Lyu}
\affiliation{National Astronomical Observatories, Chinese Academy of Sciences, A20 Datun Road, Chaoyang District, Beijing, 100101, China}
\affiliation{International centre for radio astronomy research, the university of Western Australia, 35 Stirling highway, Crawley, WA 6009, Australia}
\author[0000-0002-5927-2049]{Fanyi Meng}
\author[0000-0003-3010-7661]{Di Li}
\affiliation{Department of Astronomy, Tsinghua University, Beijing 100084, China
}
\affiliation{National Astronomical Observatories, Chinese Academy of Sciences, A20 Datun Road, Chaoyang District, Beijing, 100101, China}
\author[0000-0003-0985-6166]{Kai Zhang}
\author[0000-0002-9390-9672]{Chaowei Tsai}
\author[0000-0001-7808-3756]{Jingwen Wu}
\affiliation{National Astronomical Observatories, Chinese Academy of Sciences, A20 Datun Road, Chaoyang District, Beijing, 100101, China}
\author[0009-0007-2241-7252]{Jiachang Zhang}
\affiliation{School of Astronomy and Space Science, Nanjing University, Nanjing 210023, China}
\affiliation{Key Laboratory of Modern Astronomy and Astrophysics, Nanjing University, Nanjing 210023, China}



\begin{abstract}

Nuclear rings, prevalent in barred galaxies, are essential to understanding gas transport toward galactic nuclei. However, the peculiar nuclear ring in our neighboring galaxy M31 remains poorly understood. Here we present a comprehensive study of this multiphase gas structure, originally revealed by its dust emission, based on newly acquired CO mappings and archival spectroscopic imaging of atomic hydrogen and warm ionized gas, along with custom numerical simulations. These multi-wavelength data offer an unprecedented view of the surface mass density and kinematics of the nuclear ring, challenging the notion of it being a single coherent structure. In particular, the ring shows significant asymmetry in its azimuthal mass distribution, with neutral gas concentrated in the northeast and ionized gas prominent in the southwest. The observed off-centered and lopsided morphology disfavors an interpretation of gas streamers or resonance rings driven solely by a barred potential known to exist in M31. Furthermore, the ring's line-of-sight velocity distribution suggests circular motion in a plane inclined by $\sim 30^\circ$ relative to M31's outer disk, implying external torque likely from M32's recent close-in passage. Our hydrodynamical simulations tracking the evolution of nuclear gas of M31 influenced by both a barred potential and an oblique collision with M32, reveal the natural formation of asymmetric spiral arms several hundred Myr after the collision, which could appear ring-like under appropriate viewing angles. Therefore, we suggest that M31's nuclear gas structure, instead of being a persisting rotating ring, comprises recently formed, asymmetric spirals with a substantial tilt.

\end{abstract}



\section{Introduction} \label{sec:intro}

Galaxies are shaped by both mergers and internal secular evolution. The presence of classical bulges in early-type disk galaxies can be attributed to violent major mergers \citep{2010ApJ...715..202H} or giant clump coalescence \citep{2008ApJ...688...67E} that occurred in the distant past. On the contrary, stellar bars, which are found in about 70\% of nearby disk galaxies, are indicative of secular evolution that takes place over much longer timescales \citep{2004ARA&A..42..603K}. 
In the inner region of barred galaxies, gaseous {\it nuclear rings} and {\it nuclear spirals} are often present \citep{1993AJ....105.1344B}, which are generally thought to be the result of bar-induced inflows and subsequent concentration around the Lindblad resonance \citep{2024MNRAS.528.5742S}.
Nuclear rings and nuclear spirals can be the sites of active star formation and mediate the fueling of central supermassive black holes (SMBHs), thus directly influencing the evolution of the host galaxy \citep{2010MNRAS.407.1529H}.
Alternatively, minor mergers can also induce mass redistribution within a galaxy, leading to the formation of distorted disks and gas inflows. In extreme cases where the intruding satellite galaxy passes through the central region of the target disk galaxy, collisional rings may form as radially expanding waves in response to an impulsive gravitational perturbation \citep[e.g.][]{2010MNRAS.403.1516S, 2018MNRAS.473..585R}. 
The potential interaction of collisional rings with a preexisting nuclear ring/nuclear spiral is not well studied, let alone the effects on nuclear star formation and fueling of the central SMBH.

It has long been recognized that the central molecular zone (CMZ; \citealp{1996ARA&A..34..645M}, \citealp{2023ASPC..534...83H}), a flattened structure of predominantly dense molecular gas occupying the inner few hundred parsecs of our Milky Way, maybe the trace of a nuclear ring \citep{1991MNRAS.252..210B, 2015MNRAS.449.2421S}, although alternative geometry is also proposed, such as a two-armed nuclear spiral \citep{2017MNRAS.469.2251R} or gas streamers on a ballistic open orbit \citep{2015MNRAS.447.1059K}.
This ambiguity is largely the result of our special viewing angle toward the Galactic center, which makes the line-of-sight distance of various structures elusive. An external view is thus desired to facilitate our understanding of the formation and dynamics of CMZ-like structures and their role in the context of galaxy evolution.

Thanks to its proximity ($D \sim$ 780 kpc, where 1 arcsec $\sim$ 3.8 parsec; \citealp{2005MNRAS.356..979M, 2014AJ....148...17D}) and its many similarities with the Milky Way, the Andromeda galaxy (M31) fits this need by offering an unparalleled opportunity to scrutinize the intricate dynamical structures in a massive spiral galaxy.  In the central region of M31, well-known gaseous structures exist as nested filaments of ionized gas originally revealed in optical emission lines, in particular, H$\alpha$, which were collectively dubbed the {\it nuclear spiral} for their spiral-like morphology \citep{1985ApJ...290..136J}. The bulk of this nuclear structure was also manifested in infrared emission \citep{2006ApJ...650L..45B, 2006ApJ...638L..87G, 2009MNRAS.397..148L} and optical extinction against bulge starlight \citep{2000MNRAS.312L..29M, 2016MNRAS.459.2262D}, both unambiguously showing the presence of circumnuclear dust spatially coincident with the H$\alpha$-emitting filaments (Figure~\ref{fig:fov} and Figure~\ref{fig:flux}). 
In particular, the infrared morphology implies an elliptical ring of dusty gas, which has a projected major axis of nearly 1 kpc in length and a width of 0.2--0.3 kpc, but with a geometric center significantly offset from the center of M31 (Figure~\ref{fig:fov}). For clarity, in the following we call this dusty ring the {\it quasi-nuclear ring} or the {\it quasi-ring}, whereas the term {\it nuclear spiral} is reserved for the spiral-like features located inside the ring and to the northeast of the M31 center, clearly seen in both dust and H$\alpha$ emission.
\citet{2006Natur.443..832B} proposed that the quasi-nuclear ring is the inner one of a pair of collisional rings in response to the recent, almost head-on passage of M32, the outer ring being the so-called 10-kpc star-forming ring \citep{2006ApJ...638L..87G}.
These authors suggested that this scenario can explain the observed offset and apparent tilt of the quasi-nuclear ring. 
However, the likely presence of a stellar bar in M31 \citep{2006MNRAS.370.1499A} may also drive gas inflows toward the galactic center and produce nuclear rings and/or nuclear spirals, a case recently tested by \citet{2023arXiv231104796F}. 
Asymmetric nuclear rings and/or nuclear spirals, as observed in the CMZ, are still plausible as a consequence of unstable gas flows in a barred potential \citep{2018MNRAS.475.2383S}.

From an observational point of view, the physical properties of the quasi-nuclear ring (and the nuclear spiral) in M31 remain largely unexplored. 
This is partly due to the low-mass content of the quasi-nuclear ring/nuclear spiral, which is $\sim10^6\rm~M_\odot$ \citep{2009MNRAS.397..148L}, more than one order of magnitude lower than that of the CMZ \citep{2019MNRAS.484..964L}. 
Indeed, until not long ago it was generally thought that the center of M31 is largely devoid of neutral gas. 
Moreover, little is known about the kinematics of the quasi-nuclear ring, which otherwise promises to tell whether the ring is indeed a coherent physical structure. 
Successful detections and mapping of molecular gas via CO lines over a small portion of the quasi-nuclear ring and nuclear spiral in the past decade \citep{2011A&A...536A..52M, 2013A&A...549A..27M, 2019MNRAS.484..964L, 2019A&A...625A.148D}, the first mapping of atomic gas (via \CII and \OI emission lines in the mid-infrared) of the nuclear spiral \citep{2020ApJ...905..138L}, as well as the optical spectroscopic mapping of the circumnuclear region \citep{2018MNRAS.473.4130M, 2018A&A...611A..38O}, finally begin to unlock the full potential of M31 to advance our understanding of the formation and evolution of nuclear dynamical structures.

Here we present a comprehensive analysis of the quasi-nuclear ring in M31 based on our newly obtained CO observations, in combination with multi-wavelength data and custom numerical simulations. 
%Our results show that the distribution of gas in the M31 center is not only asymmetric, but also exhibits a strong variation in phases. 
Section \ref{sec: obs} describes the data sets used in this study, including our new IRAM CO(1-0) and JCMT CO(3-2) observations focusing on the quasi-nuclear ring, recent VLA and FAST HI surveys of the entire M31 disk, and CFHT/SITELLE IFU observations covering the nuclear region of M31. Data reduction procedures are also described in detail. In Section \ref{sec: results} we present the properties (surface mass density and line-of-sight velocity) of the neutral and ionized gas across the quasi-nuclear ring, which establish a previously overlooked lopsidedness of the ring and in turn challenges it being a coherent structure. 
In Section~\ref{sec:discussion}, we discuss the possible origins of the quasi-nuclear ring, assisted with custom numerical simulations introduced in Section~\ref{sec:sim}. We favor a hybrid scenario in which a bar-driven inflow perturbed by a recent close-in passage of M32 is responsible for the observed ring-like structure.
A summary of our study is given in Section \ref{sec: sum}. 

\section{Observation and data reduction} \label{sec: obs}

\begin{figure}
    \centering
    \plotone{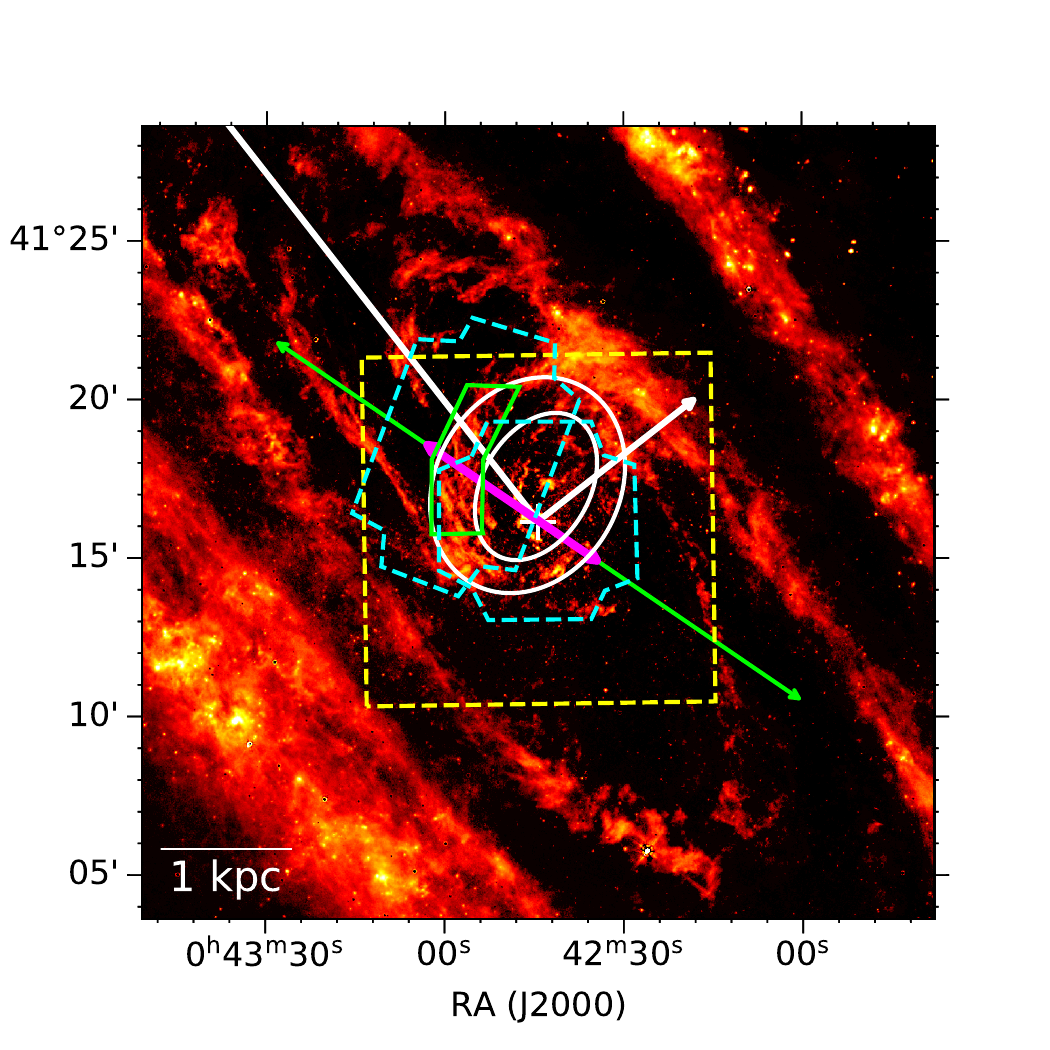}
    \caption{A {\it Spitzer}/IRAC 8 $\mu$m image of M31 \citep{2006ApJ...650L..45B}, overlaid with the field-of-view of IRAM 30m CO(1-0) (solid green polygon), JCMT CO(3-2) (dashed cyan polygons; see text for details), and CFHT/SITELLE IFU (dashed yellow box). The M31 center is marked with a white cross. An off-centered ``quasi-nuclear ring'' is outlined with two white ellipses. The position of the bar derived from the made-to-measure model \citep{2017MNRAS.466.4279B, 2018MNRAS.481.3210B} is indicated with a pair of green arrows, with a position angle of 55.7$^\circ$ and a half length of 4 kpc. The corotation radius (6.5 kpc) of the bar is indicated with two white arrows along the major and minor axes of the disk, while the inner inner Lindblad resonance (1.0 kpc) and outer inner Lindblad resonance (1.8 kpc) radii are marked with the two opposite magenta arrows along the bar major axis. 
    }
    \label{fig:fov}
\end{figure}

We have made use of our newly obtained IRAM 30m and JCMT CO observations and publicly available ancillary multi-wavelength observations to help reveal the origin of the quasi-nuclear ring. Figure \ref{fig:fov} shows the $Spitzer$/IRAC 8 $\mu$m image \citep{2006ApJ...650L..45B}, with the stellar bulge subtracted and tracing emission from polycyclic aromatic hydrocarbons molecules, which covers the main part of the M31 disk and highlights the features known as the quasi-nuclear ring and the nuclear spiral (located inside the ring).
Overlaid on this image is the field-of-view (FoV) of the multi-wavelength data used in this study. The position of the bar and related resonance radii are also indicated (\citealp{2018MNRAS.481.3210B}, see Section \ref{subsec:bar} for details). We also used data from existing CO and HI surveys of M31. The basic information of various datasets and the reduction procedures are described below.

%%% JCMT observation
\subsection{JCMT CO(3-2) observations}\label{sec:obs-CO32}

Observations of the CO(3-2) line were conducted using the JCMT telescope (Program ID M19BP006; PI: Z. Li). The Heterodyne Array Receiver Program (HARP) was utilized, specifically the 4 $\times$ 4 receptor array, which has a fast scanning capability \citep{2009MNRAS.399.1026B}. The observations were made between October and December 2019, with a total integration time of 20.5 hours under band 2 weather. The target region for the observations was the northeastern half of the quasi-nuclear ring, which, centered around 00:42:54.8153, +41:19:03.8424 (J2000), is the brightest part of the quasi-nuclear ring in terms of CO and dust emission (Figure \ref{fig:fov}). For clarity, we call the northeastern segment the NE arc in the following, to distinguish from the southwestern segment that is less prominent in neutral gas (hereafter the SW segment). To minimize uncertainties from the sky and system, the position-switch basket weave mode was employed. This mode involves scanning the field of interest horizontally and perpendicularly using the 14 working receptors of HARP, excluding H14 and H16. The scanning steps were set in a 1/4 array (7.3$''$) with a separation of 30$''$, achieving Nyquist sampling. The resulting raster map covered a region of approximately 2 $\times$ 4 arcmin$^2$ (equivalent to approximately 456 $\times$ 912 pc$^2$) on the ring. Furthermore, it also covered an undersampled outer region with a width of approximately 2$'$ (dashed cyan polygon in Figure \ref{fig:fov}). All observations had a bandwidth of 1000 MHz and a spectral resolution of 0.488 MHz, corresponding to a velocity resolution of 0.423 km s$^{-1}$. %A detailed log of the observations can be found in Table 1.

The standard procedure for JCMT heterodyne observations was followed for data reduction, employing the ORAC-DR pipeline \citep{2015A&C.....9...40J}, which is based on the STARLINK software \citep{2014ASPC..485..391C}. The pipeline performed automatic tasks, such as flagging spikes and bad detectors, subtracting linear baselines, and evaluating data quality. Using the spectra, we created a position-position-velocity data cube with a spatial resolution of 15$''$, which is approximately the beam size at 346 GHz. The baseline RMS varied due to the varying integration time on the map, with a typical value of 3.5 mK (in 10 km s$^{-1}$ channels) near the center of the field, gradually increasing to approximately 23 mK in the outer region. This sensitivity is consistent with our previous JCMT CO(3-2) mapping that covered the inner $2' \times 2'$ region (\citealp{2019MNRAS.484..964L}, with an undersampled outer region shown with dashed cyan polygon roughly centered on the M31 center in Figure \ref{fig:fov}). 
% Proposal: 16.3 mK TA*. 
To achieve a more holistic mapping of the nuclear region, we combined the two datasets using the WCSMOSAIC task in STARLINK.

To obtain the moment maps, we followed the methods described by \citet{2012MNRAS.424.3050W}. The process can be summarized as follows. To convert from antenna temperature to main-beam temperature, we divided each spectrum by the main-beam efficiency ($\eta\rm_{MB}$) of 0.64. Since the signal-to-noise ratio (S/N) varies across the FoV due to factors such as system temperatures, integration times, and unstable receptors, we divided the original data cubes by the noise maps generated using the line-free channels of each of the two fields. This allowed us to obtain the S/N cube. Using pixels with S/N greater than three times the local RMS, we then created the integrated intensity map (m0 map) and the intensity-weighted line-of-sight velocity map (m1 map). 
The line-of-sight velocity was corrected to the local standard of rest (LSR).
The results are presented in Figures \ref{fig:flux} and \ref{fig:v0}. To convert the integrated intensity of CO(3-2) to the molecular gas surface density, a line ratio of CO(3-2)/CO(1-0) of 0.8 is adopted, which is typical of CO emission in this region \citep{2019MNRAS.484..964L}. The conversion factor $X\rm_{CO(1-0)} = 2 \times 10^{20}~cm^{-2}~(K~km~s^{-1})^{-1}$, conventional for the Milky Way molecular gas \citep{2013ARA&A..51..207B}, is then adopted to convert the intensity of CO to the molecular hydrogen column density and then to the surface mass density of molecular gas ($\Sigma_{\rm H_2}$, in units of $\rm M_\odot~pc^{-2}$).

%%% IRAM observation
\subsection{IRAM 30m CO(1-0) observations}\label{sec:obs-CO10}

CO(1-0) observations were carried out with the IRAM 30m telescope in July 2019 (Program ID 058-19 and 185-19; PI: Z. Li), with a total integration time of 20 hours covering the NE arc of the quasi-nuclear ring. The FoV is a mosaic of two $1.67'\times 2.33'$ regions (the second one has a position angle of 335$^\circ$), covering an area of 7.8 arcmin$^2$ in total. The two regions are designed to better match the curvature of the ring, as shown in Figure \ref{fig:fov} (solid green polygon). The observation mode was the on-the-fly (OTF) PSW mode to map the targeted region coherently. The sensitivity of the final data is 5.5 mK in a 10.4 km s$^{-1}$ channel. 

We used the CLASS tool in the GILDAS software package\footnote{http://www.iram.fr/IRAMFR/GILDAS} \citep{2005sf2a.conf..721P}  developed by IRAM to examine and process the spectra. First, we checked the quality of individual spectra to ensure there were no spikes or bad channels. Then we performed platform correction on each FTS spectrum using the script FtsPlatformingCorrection5.class\footnote{http://www.iram.es/IRAMES/mainWiki/AstronomerOfDutyChecklist} by subtracting baselines from the individual FTS units. %After the correction, we shifted the central frequency to match the redshift of each source and used a line-free region with a total width of 5000 km s$^{-1}$ to subtract a first-order baseline. 
The antenna temperature $T\rm_A^*$ was converted to the brightness temperature of the main beam by $T\rm_{mb}$ = $T\rm_A^*$ $F\rm_{eff}$/$\eta\rm_{mb}$, with main beam efficiency $\eta\rm_{mb}$ = 0.78 and the forward beam efficiency $F\rm_{eff}$ = 0.94 at 115 GHz\footnote{http://www.iram.es/IRAMES/mainWiki/Iram30mEfficiencies}.

Furthermore, we obtained data from the IRAM 30m CO(1-0) survey conducted by \citet{2006A&A...453..459N}. This survey covered the entire M31 disk with a FoV of $2^\circ \times 0.5^\circ$ and had the same angular resolution as our own observation. The average RMS noise of the CO(1-0) m0 map is 0.35 K km s$^{-1}$ throughout the disk and is $\sim$0.1 K km s$^{-1}$ in the central region. We used the imregrid task in CASA \citep{2007ASPC..376..127M} to truncate the CO(1-0) image to match the fields of our observations. Finally, we merge the m0 map with our observations of the quasi-nuclear ring using the WCSMOSAIC task in STARLINK. This allowed us to achieve a more complete measurement of the quasi-nuclear ring in CO(1-0). Moment maps were obtained using the same method as for the CO(3-2) data. The integrated intensity was then converted to molecular hydrogen column density using 
%the conventional conversion factor 
$X\rm_{CO} = 2 \times 10^{20}~cm^{-2}~(K~km~s^{-1})^{-1}$ %\citep{2013ARA&A..51..207B} 
and then to the surface mass density in units of $\rm M_\odot~pc^{-2}$. %[wouldn't it be better to move these sentences to the next paragraph?] 
A more in-depth study of CO(1-0) and CO(3-2) emission from the circumnuclear region is deferred to future work.

\subsection{Optical IFU observations}\label{sec:obs-IFU}

We also utilized integral-field unit (IFU) observations with the SITELLE imaging Fourier transform spectrograph installed on the Canada-France-Hawaii Telescope (CFHT) \citep{2019MNRAS.485.3930D}. The IFU data were retrieved from the Canadian Astronomical Data Center (CADC\footnote{https://www.cadc-ccda.hia-iha.nrc-cnrc.gc.ca/AdvancedSearch}). The observations (Proposal ID: 16BC19, \citealp{2018MNRAS.473.4130M}) cover an $11' \times 11'$ region centered on the nucleus of M31, using filters SN2 (480--520 nm) and SN3 (647--685 nm) to characterize emission line stellar objects (e.g., planetary nebulae) as well as circumnuclear ionized gas.
%properties of the nuclear ring and surrounding regions. 
The SN2 filter includes H$\beta$ and the [OIII]$\lambda\lambda$4959,5007 doublets, while the SN3 encompasses H$\alpha$ and the [NII]$\lambda\lambda$6548,6584 and [SII]$\lambda\lambda$6716,6731 doublets. 
The observations have a spatial sampling of 0.32$''$ and a seeing of $\sim 1''$. The spectral resolution is 5000 (full width at half maximum of 1.3$\rm~\AA$) for the data with the SN3 filter, which we use primarily here. More details on the observations can be found in \citet{2018MNRAS.473.4130M}.

The data processing followed the standardized pipeline of the SITELLE instrument using the ORBS software \citep{2012SPIE.8451E..3KM, 2015ASPC..495..327M}. The main steps of the pipeline include spatial calibration, wavelength calibration, and flux calibration, as detailed in \citet{2018MNRAS.473.4130M}. The wavelength was corrected using sky lines as described in \citet{2016sf2a.conf...23M}. 
After skyline correction, the line-of-sight velocity was converted to the LSR.

In this study, we focus on the H$\alpha$ emission line, since it is the most representative line for the ionized gas, and its morphology and kinematics trace the ionized gas well. A comprehensive analysis of the other emission lines and the stellar populations will be presented in future work. 
A two-step procedure was applied to obtain a satisfactory fit for the H$\alpha$ line. 
The first step involves masking out the emission lines and fitting the stellar continuum in SN2 to determine the stellar velocity field since SN2 contains abundant stellar absorption lines as well as an on-average higher S/N. 
The native SITELLE spaxels were first binned by a factor of 4 to increase the S/N. Even after this binning, many pixels still did not show clear emission lines. Therefore, the Voronoi tessellation binning \citep{2003MNRAS.342..345C} was further employed based on the full-filter stellar continuum measured in the SN2, in order to achieve an S/N of 2000 for each bin, which resulted in 1655 bins in total.
Each of the 1655 bins was then fitted for the continuum using the penalized pixel-fixing method (pPXF, \citealp{2004PASP..116..138C}). The E-MILES stellar population template \citep{2016MNRAS.463.3409V} was adopted for fitting the stellar continuum. 

In the second step, the stellar velocity field derived from the SN2 data was used as a fixed parameter to fit the stellar continuum in SN3, based on the same Voronoi tessellation. The H$\alpha$ and the [N\,{\sc ii}] and [S\,{\sc ii}] doublets were fitted simultaneously in this step, each with a Gaussian profile. Line-of-sight velocity and velocity dispersion were free parameters and assumed to be the same among the lines.  
A sinc function with a width of 1.3 $\rm \AA$, specific to the Fourier transform spectrograph, was adopted to account for the instrumental broadening. 
We note that in some regions the H$\alpha$ exhibits a double-peaked line profile, suggesting the presence of two velocity components in these regions, which was previously noted by \citet{2018A&A...611A..38O} based on the VIRUS-W IFU observation. A double Gaussian profile, with a fixed velocity dispersion at 40 km~s$^{-1}$ (the averaged value inferred from the single Gaussian model), was used to refit the 1655 bins, and an F-test was conducted to evaluate the likelihood that the double Gaussian profile is a better model for the data. We found that only $\sim$20\% of the bins show a statistically significant double component, mostly in regions outside the quasi-nuclear ring and nuclear spiral. Therefore, we decide that a single Gaussian fit is sufficient for our present purpose of characterizing the spatial distribution and kinematics of the H$\alpha$ line along the ring.

The resultant H$\alpha$ intensity map is shown in Figure \ref{fig:optical}, with a typical RMS of $2\times 10^{-17}\rm~erg~s^{-1}~cm^{-2}~arcsec^{-2}$ in the quasi-nuclear ring, after correcting for the Galactic extinction $A\rm_V = 0.17$ and a typical intrinsic extinction of the M31 nuclear region, $A\rm_V = 0.014$ \citep{2016MNRAS.459.2262D}.
The surface mass density of the ionized gas can be estimated from the flux of H$\alpha$, using the equation $\Sigma_{\rm HII} = \mu \it m\rm_p \it I\rm_{H\alpha}/\it \Lambda n\rm_e$. Here $I\rm_{H\alpha}$ is the H$\alpha$ intensity, %$D$ is the distance of M31, 
$\mu=1.4$ is the mass per H atom, $m\rm_p$ the proton mass, $n\rm_e$ the electron density and $\Lambda = 3.56 \times 10^{-25}\rm ~erg~cm^3~s^{-1}$ is the effective volume emissivity for case B recombination at $T = 10^4 \rm ~K$ \citep{1989agna.book.....O}. Here, we adopted a typical electron density of $\sim$400 cm$^{-3}$ derived from the line ratio of the [S\,{\sc ii}] doublet across the nuclear region (Z. Li et al. in prep.). The resultant surface density of the ionized gas is compared to that of the other phases of the quasi-nuclear ring in Figure~\ref{fig:mass_dist}.

\subsection{HI observations}\label{sec:obs-HI}

We retrieved archival Karl G. Jansky Very Large Array (VLA) HI observations covering the entire disk of M31 (Program ID: 14A-235; PI: Adam Leroy) in D-configuration with a beam size of $\sim 45''$. These HI observations together resulted in a 51-field mosaic in the L-band, with
1.95 kHz spectral channels, corresponding to a velocity resolution of 0.42 km s$^{-1}$. 
The data reduction was performed using the standard VLA pipeline in CASA v5.4.1, with some adjustments made specifically for spectral line data \citep[see details in][]{2021MNRAS.504.1801K}. For each track, the pipeline was run iteratively, with manual flagging applied when necessary. In most cases, two pipeline runs were required to obtain well-calibrated data. The typical RMS noise is derived from the emission-free region on the m0 map, which is 0.13 Jy beam$^{-1}\rm ~km~s^{-1}$. 

To recover diffuse HI emissions in the circumnuclear region, we also made use of Five-hundred-meter Aperture Spherical radio Telescope (FAST) HI observations covering the whole disk of M31 (K. Zhang et al. in prep.). The observations were made with the FAST L-band 19-feed receiver in drift scan mode. The scans were separated by 21.9$'$ to achieve super-Nyquist sampling \citep{2018IMMag..19..112L, 2019SCPMA..6259506Z}.
Data reduction follows standard procedures, including electronic gain calibration, bandpass calibration, baseline fitting and ripple removal, and flux calibration. The reduced cube is a combination of narrow and wideband data, with a velocity coverage from $-1000$ to 500 km s$^{-1}$, a channel width of 1.6 km s$^{-1}$, and a sensitivity of 0.01 Jy beam$^{-1}$.

The FAST and VLA data are then combined using the J-COMB algorithm developed by \citet{2022SCPMA..6599511J} to achieve high sensitivity and resolution (the beam size of the combined map is 1$'$, similar to the resolution of VLA). Unlike the CO and H$\alpha$ observations, which predominantly exhibited characteristics compatible with single Gaussian fitting, the HI emission exhibits apparent multi-velocity components due to its diffuse nature and the lower angular resolution of the observations. In particular, up to five HI components have been detected in M31 \citep{2009ApJ...705.1395C}. Broad velocity components at around $-300$ km s$^{-1}$ are recovered in the central region of M31 with sensitive FAST observations, which are potentially background/foreground emissions arising from the thickened outer disk projected to a nearly edge-on view \citep{2010A&A...511A..89C, 2011A&A...536A..52M}. Therefore, a multi-Gaussian fitting approach was adopted (\citealp{2023MNRAS.524.1169L}, see Appendix \ref{sec:HI_spec} for details) to separate the background/foreground component from the distinct nuclear component. Subsequently, %on the basis of position-velocity (PV) diagrams, 
the nuclear component was extracted to generate the m0 and m1 maps. After subtraction, an arc-like feature appeared on the NE side, as shown in Figure \ref{fig:flux}, co-spatial with the NE arc prominent in dust emission. We further clarify that we did not extract this feature by strictly defining a quasi-ring region. Instead, the feature emerged naturally from our analysis of the entire nuclear region, where we separated the outer disc from other kinematic components.
%The m0 and m1 maps were derived using the same method applied to the CO data. 
The integrated intensity was then converted to atomic hydrogen column density using $N\rm_H = 1.823\times 10^{18} \times \it T\rm_B d\it v~ \rm cm^{-2}$, under the reasonable assumption that the HI gas is optically thin in this region.

\section{The quasi-nuclear ring in multi-phase}
\label{sec: results}

This section presents a multi-wavelength view of the quasi-nuclear ring traced by different phases of the medium (H$_2$, HI, HII, and dust). The surface density map (Section~\ref{subsec:surfacemap}) and velocity field (Section~\ref{subsec:kinematics}) of different phases are compared, and the azimuthal distribution of surface density and velocity along the ring are also investigated, which reveal significant azimuthal variation in all phases. We note that only HI exhibits a dominant component around -300 $\rm km~s^{-1}$, likely originating from the thickened outer HI disk and projected into the center. To isolate the nuclear emission, we subtracted this diffuse outer disk component via multi-Gaussian fitting (see Appendix \ref{sec:HI_spec} for details). In contrast, CO and H$\alpha$ are unaffected by this contamination, allowing their integrated intensity and velocity field maps to be computed directly from their full spectral range. All subsequent analysis is based on these processed maps.

\subsection{Surface density distribution of different components}
\label{subsec:surfacemap}

\begin{figure}
    \centering
    \includegraphics[width=\linewidth]{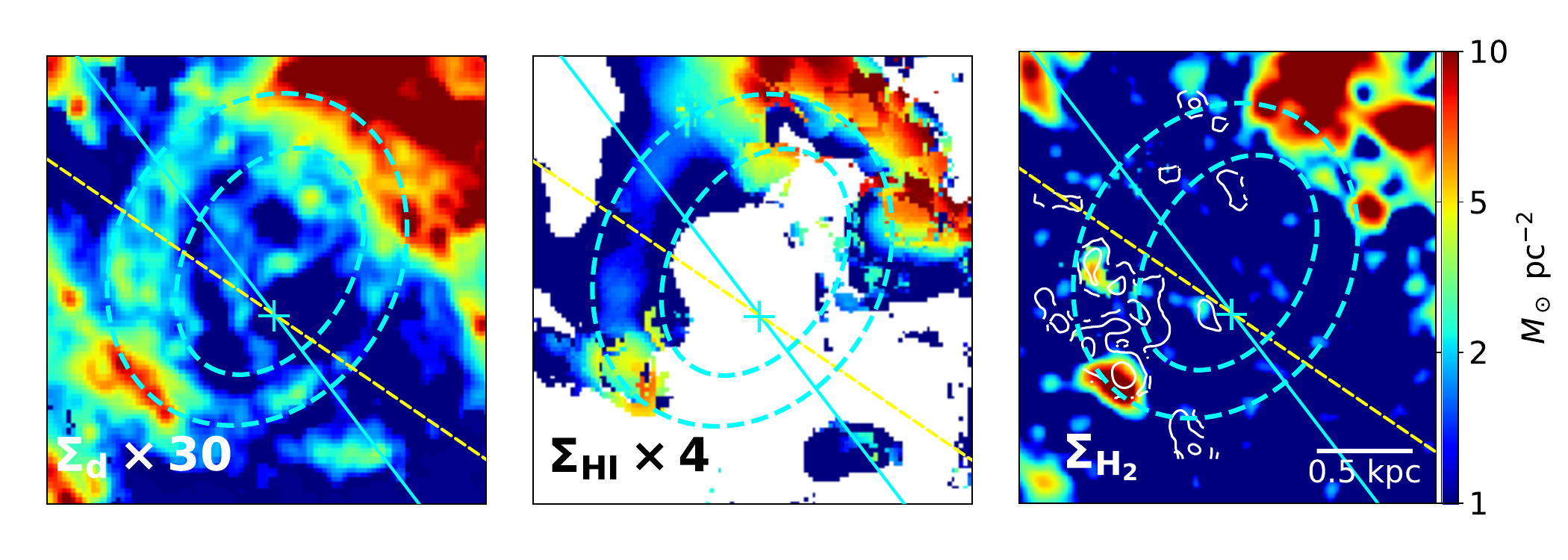}
    \caption{From left to right: surface mass density maps of dust \citep{2012MNRAS.426..892G}, atomic hydrogen (F. Meng et al. in prep.) and molecular hydrogen (this work). The dust and HI surface density are multiplied by a factor of 30 and 4, respectively, to match the scale of the H$_2$ gas better. A velocity component around $-300$ km s$^{-1}$ lying in the foreground/background has been subtracted from the HI data to obtain the atomic hydrogen surface density of the quasi-nuclear ring (see Appendix \ref{sec:HI_spec} for details). The molecular hydrogen surface density is derived from IRAM 30m CO(1-0) observations, while the white contours show the surface density of 1, 2, 5, and 10 M$_\odot$ pc$^{-2}$ derived from JCMT CO(3-2) observations. 
    Here a constant CO(3-2)/CO(1-0) line ratio of 0.8 typical in the nuclear region is adopted \citep{2019MNRAS.484..964L}. The dashed cyan ellipses outline the quasi-nuclear ring that is prominent in dust emission. The cyan cross marks the center of M31. The cyan line marks the major axis of the M31 disk, while the major axis of the bar that deviates from the disk major axis by 17.7$^\circ$ \citep{2017MNRAS.466.4279B} is indicated with a dashed yellow line.}
    \label{fig:flux}
\end{figure}

\begin{figure}[ht!]
%\plotone{samplefig.png}
\plotone{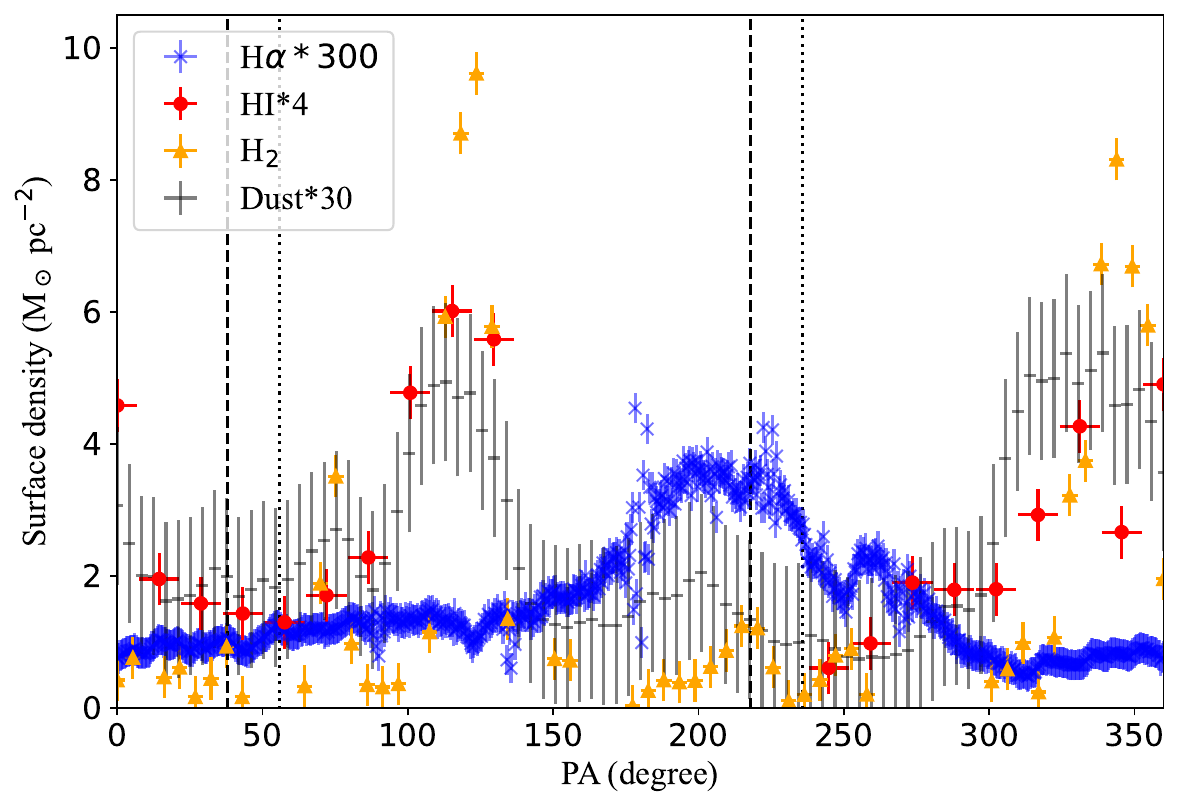}
\caption{The surface density distribution as a function of the position angle (measured east from the north with respect to the M31 center) extracted from the quasi-nuclear ring region defined with the dashed ellipses in Figure \ref{fig:flux}. For clarity, the distributions of dust, HI, and H$\alpha$ are normalized by a factor denoted in the insert. The ring region is divided into azimuthal bins according to the resolution of different images, and the mean value within each bin is adopted as the representative surface density. The errorbar on the y-axis represents the typical RMS of the data. The HI profile is incomplete due to the limited sensitivity of the HI data for some part of the ring region. The major axis of the disk is marked by a pair of vertical dashed lines, while that of the bar is shown with a pair of dotted lines. Significant azimuthal variations are present in all phases and between the neutral and ionized phases. 
\label{fig:mass_dist}}
\end{figure}

The surface density maps of dust, atomic hydrogen, and molecular hydrogen are shown in Figure \ref{fig:flux}. The dust surface density map is retrieved from \cite{2012MNRAS.426..892G}. This map is derived by fitting images of $Herschel$ PACS 100, 160 $\mu$m and SPIRE 250 $\mu$m images with a modified blackbody, which is sensitive to a dust mass surface density of 0.04 $\rm M_\odot~pc^{-2}$. On the other hand, the surface densities of HI and H$_2$ are sensitive to $\sim 0.3\rm~and~1 ~M_\odot~pc^{-2}$, respectively. The gas-to-dust ratio averaged over the quasi-nuclear ring is $17.4 \pm 13.8$ (with a maximum of 50 at the southeastern portion), much lower than that in the disk ($\sim$100, \citealp{2014ApJ...780..172D}), while roughly consistent with the value found in the nuclear region ($\sim 36\pm 6$, \citealp{2014ApJ...780..172D, 2019MNRAS.484..964L}). 
The quasi-nuclear ring, originally defined through dust emission, is visible on the dust map, with a surface density greater than $\sim 0.06\rm~M_\odot~pc^{-2}$ for the most part. Still, notably the quasi-ring is wider in the NE arc, with a width of $\sim$300 pc, compared to $\sim$100 pc in the south. %brighter in its northeastern segment, with a surface density $\sim 0.1-0.15\rm~M_\odot~pc^{-2}$.  
The ring is even more asymmetric in the neutral gas than in the dust. HI emission displays an incomplete ring morphology, similar to that found by \citet{2011A&A...536A..52M} after subtracting the component around the systemic velocity of M31 from the WSRT HI datacube \citep{2009ApJ...695..937B}, although they did not specifically isolate the component belonging to the ring. HI is notably bright in the southeastern and northwestern portions of the ring. These positions coincide with the two apparent intersections between the ring and the outer spiral arms, where the gas density is presumably enhanced to produce the two peaks in Figure \ref{fig:mass_dist}.
Similarly, the molecular gas traced by the integrated intensity of CO(1-0) and CO(3-2) emission is significantly stronger in the southeastern and northwestern regions, consistent with that of the HI distribution. The NE arc in dust emission is evident, but relatively faint in HI emission, while several bright CO clumps are identified in the lower half of this segment. 
This is the first time that the NE arc of the quasi-nuclear ring is unambiguously detected in both atomic and molecular gas phases, supporting the notion of a genuine gas structure. 
%strengthening the case of a coherent structure. 
On the other hand, the SW segment of the ring, which is weaker in dust emission, also shows little or no emission of HI or CO, casting doubt on the reality of a complete gas ring.

\begin{figure}
    \centering
    \includegraphics[width=\linewidth]{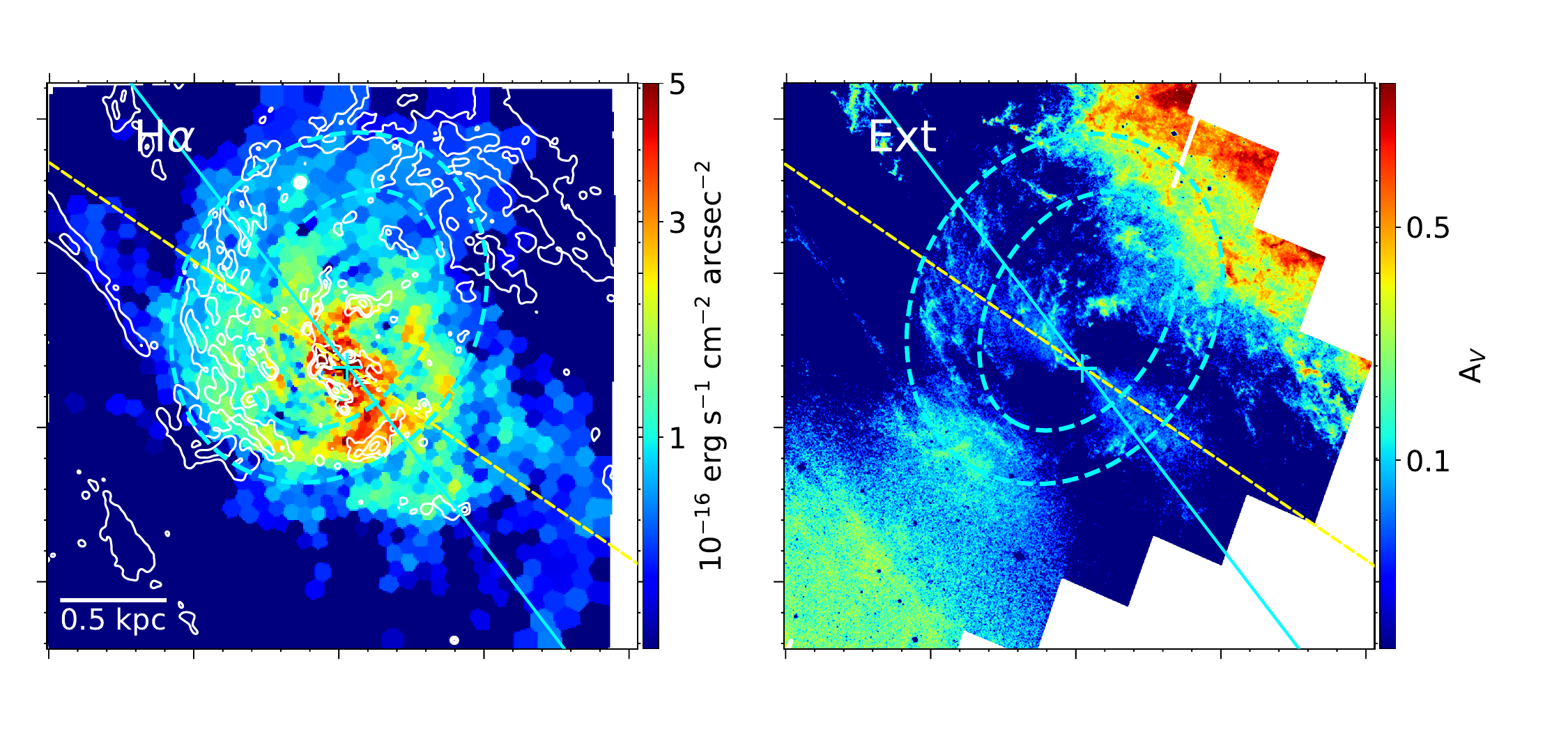}
    \caption{$Left$: CFHT/SITELLE H$\alpha$ intensity distribution overlaid with contours from the $Spitzer$/IRAC 8 $\mu$m image shown in Figure \ref{fig:fov}. The northern half of the ring is weak in ionized gas emission. $Right$: The extinction map derived from HST/WFC3 F475W images from the PHAT survey \citep{2012ApJS..200...18D}. The dashed cyan ellipses outline the quasi-nuclear ring. 
    The cyan cross marks the center of M31. 
    The dashed cyan line marks the major axis of the M31 disk, while the major axis of the bar that deviates from the disk major axis by 17.7$^\circ$ \citep{2017MNRAS.466.4279B} is indicated with a dashed yellow line. }
    \label{fig:optical}
\end{figure}

Further complicating the picture, the ionized gas exhibits a distinct morphology, as manifested in the H$\alpha$ intensity map presented in the left panel of Figure \ref{fig:optical}. The ionized gas is characterized by bright spiral-like filaments across the central kpc. Remarkably, the H$\alpha$ emission is bright in the southern portion of the ring but appears much fainter and fuzzier in the northern part, almost exactly opposite to the situation of the dust and neutral gas.
For comparison, a high-resolution dust extinction map derived from HST images from the Panchromatic Hubble Andromeda Treasury Survey (PHAT, \citealp{2012ApJS..200...18D}) is also presented (right panel of Figure \ref{fig:optical}). Specifically, the extinction map is derived by modeling and subtracting the bulge starlight in the F475W image with {\sc galfit} software \citep{Peng_2002, Peng_2010} using the same method as described in \citet{2022ApJ...928..111L}. The typical value of apparent extinction $A\rm_V$ across the ring region is $\lesssim 0.1$, which is consistent with a mean column density $\lesssim 10^{20}\rm ~cm^{-2}$ derived from the HI/CO maps,
%ring/northeastern arm is $\lesssim 0.5/0.1$, 
indicating little effect on the morphology of the optical emission lines. 
Remarkable feather-like features in the NE arc are revealed in the high-resolution extinction map (also appreciable in the dust emission shown in Figure~\ref{fig:fov}), which protrude roughly perpendicularly from the segment and bend slightly northwestward.
%suggest the presence of turbulence in this region. 
These features are reminiscent of patterns generated by the wiggle instability often seen in simulations of spiral shocks \citep{2004MNRAS.349..270W} and the so-called off-axis shocks induced in a bar potential \citep[][also known as dust lanes]{2012ApJ...747...60K}, which is understood to be due to vorticity generation in curved shocks \citep{2014ApJ...789...68K}.

To further quantify the aforementioned trend of surface density variations in the different components of the quasi-nuclear ring, we calculate the azimuthal distribution of surface density of each phase. To achieve this, the ring region defined by the two ellipses in Figure \ref{fig:flux} is isolated. 
This region is then divided into a sequence of azimuthal bins as a function of the position angle (PA; north as zero and increasing counterclockwise with respect to the M31 center). 
The size of the azimuthal bins varies according to the original image resolution of a given phase, and the representative value for each bin is determined as the mean value of spaxels included in the bin for a given phase.
The resulting azimuthal surface density distributions are shown in Figure \ref{fig:mass_dist}. Significant azimuthal variation is present, by a factor of up to 10, in both the dust and gas surface densities. The two highest peaks at PA $\sim$ 130$^\circ$ and 340$^\circ$ presented in the dust and neutral gas distributions correspond to the positions where the ring apparently connects with the outer spiral arms, which are close to the minor axis of the galactic disk. We stress that the conclusion of a significant azimuthal variation in the neutral gas holds even if these two peaks were not taken into account. The surface density of molecular hydrogen averaged over the two peaks is $\sim 4$ times higher than that of atomic hydrogen, while the molecular and atomic gas mass summed over the quasi-nuclear ring region defined here is $\sim 2\times 10^6 \rm~M_\odot$ and $\sim 5\times 10^5 \rm~M_\odot$, respectively. 
These values are broadly consistent with the atomic-to-total gas fraction of 20--40 percent in the central region estimated from the IRAM 30m CO(1-0) survey \citep{2006A&A...453..459N}. This indicates that the quasi-nuclear ring is predominantly molecular gas, similar to the situation of the CMZ. 

On the other hand, the surface density of the ionized gas has a broad peak at PA $\sim 180^\circ - 220^\circ$, close to the direction of the disk major axis (marked by a pair of vertical dashed lines), where the surface density of the neutral gas is generally low. Conversely, the surface density of the ionized gas is generally low over position angles where the neutral gas peaks. These trends are consistent with Figures~\ref{fig:flux} and \ref{fig:optical}. We note that the surface density of the ionized gas is dependent on the assumptions on electron density and case B recombination, and the surface density of molecular gas is dependent on the $X\rm_{CO}$ factor, which are of substantial uncertainties \citep{2013ARA&A..51..207B}. Nevertheless, the qualitative difference in the azimuthal distribution between the neutral and ionized gas phases should be robust.

\subsection{Gas kinematics \label{subsec:kinematics}}

\begin{figure}
        \centering
        \includegraphics[width=\linewidth]{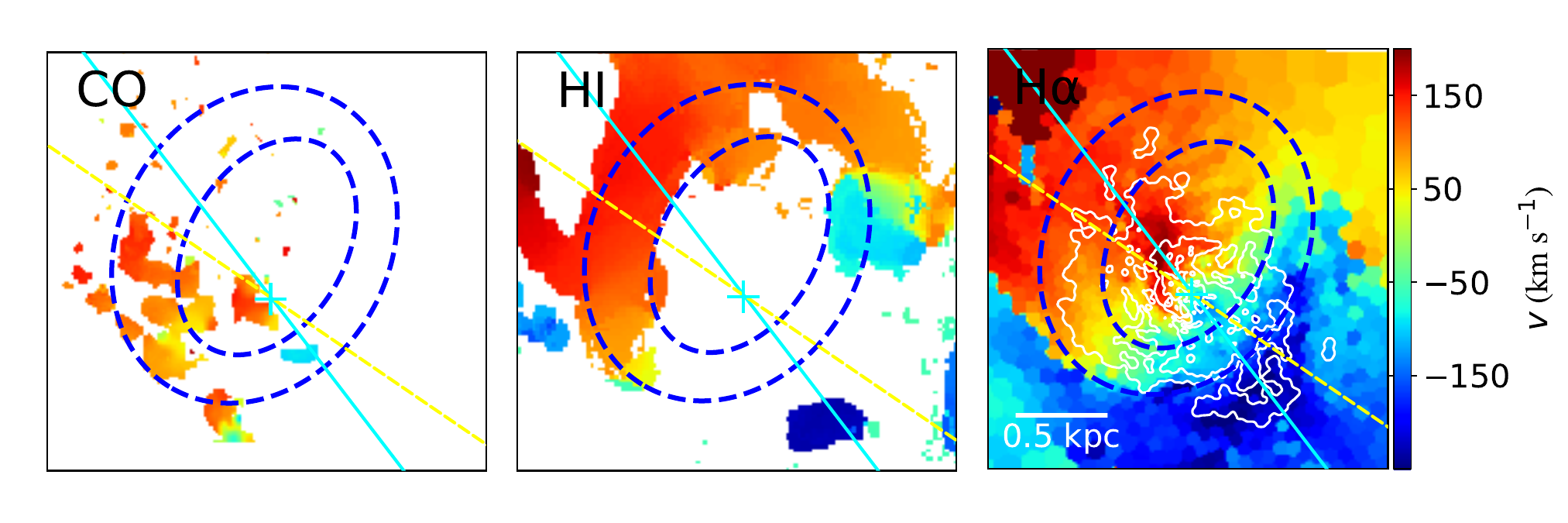}
        \caption{From left to right: Line-of-sight velocity field of CO(3-2), HI, and H$\alpha$. The H$\alpha$ velocity field is overlaid with the contours from the H$\alpha$ intensity image presented in Figure \ref{fig:optical}, with levels 5, 10, 20 times the RMS level.} CO(3-2) is chosen as the tracer of the velocity field of the molecular gas because it has higher resolution and better sensitivity than CO(1-0). The HI velocity field of the ring component is separated using the FMG tool (see Appendix \ref{sec:HI_spec} for details), with the $-300$ km s$^{-1}$ diffuse component subtracted. The velocity field of the three gas phases follows a similar pattern, indicating a clockwise rotational motion when viewing from the north pole, consistent with the bulk rotation of the galactic disk. The velocities have been corrected for the systemic velocity of M31 ($-300\rm~km~s^{-1})$. 
        \label{fig:v0}
\end{figure}

The line-of-sight velocity distribution of various gas phases, including CO, HI, and H$\alpha$, is shown in Figure \ref{fig:v0}. The velocity field of molecular gas is derived from CO(3-2) instead of CO(1-0) since the sensitivity of the former is higher and the resultant velocity field is smoother. Although with a smaller FoV, the CO(3-2) observation still covers a large fraction of the quasi-nuclear ring, in particular the NE arc.
The HI velocity field of the ring component shows a clear velocity gradient and is in rough agreement with that of CO and H$\alpha$. 
All three phases follow an overall clockwise rotation pattern when viewing from the north pole, that is, redshifted on the NE side and blueshifted on the SW side, consistent with the galactic rotation. The rotational direction and orientation of the feather-like substructures in the NE arc, as illustrated in Figure \ref{fig:optical}, collectively indicate an inspiraling pattern in this region.
The highest velocity in the ring region is approximately $\pm$150 km s$^{-1}$ (after correcting for a systemic velocity of $-300$ km s$^{-1}$) and is roughly aligned with the disk major axis.

\begin{figure}[ht!]
%\plotone{samplefig.png}
\plotone{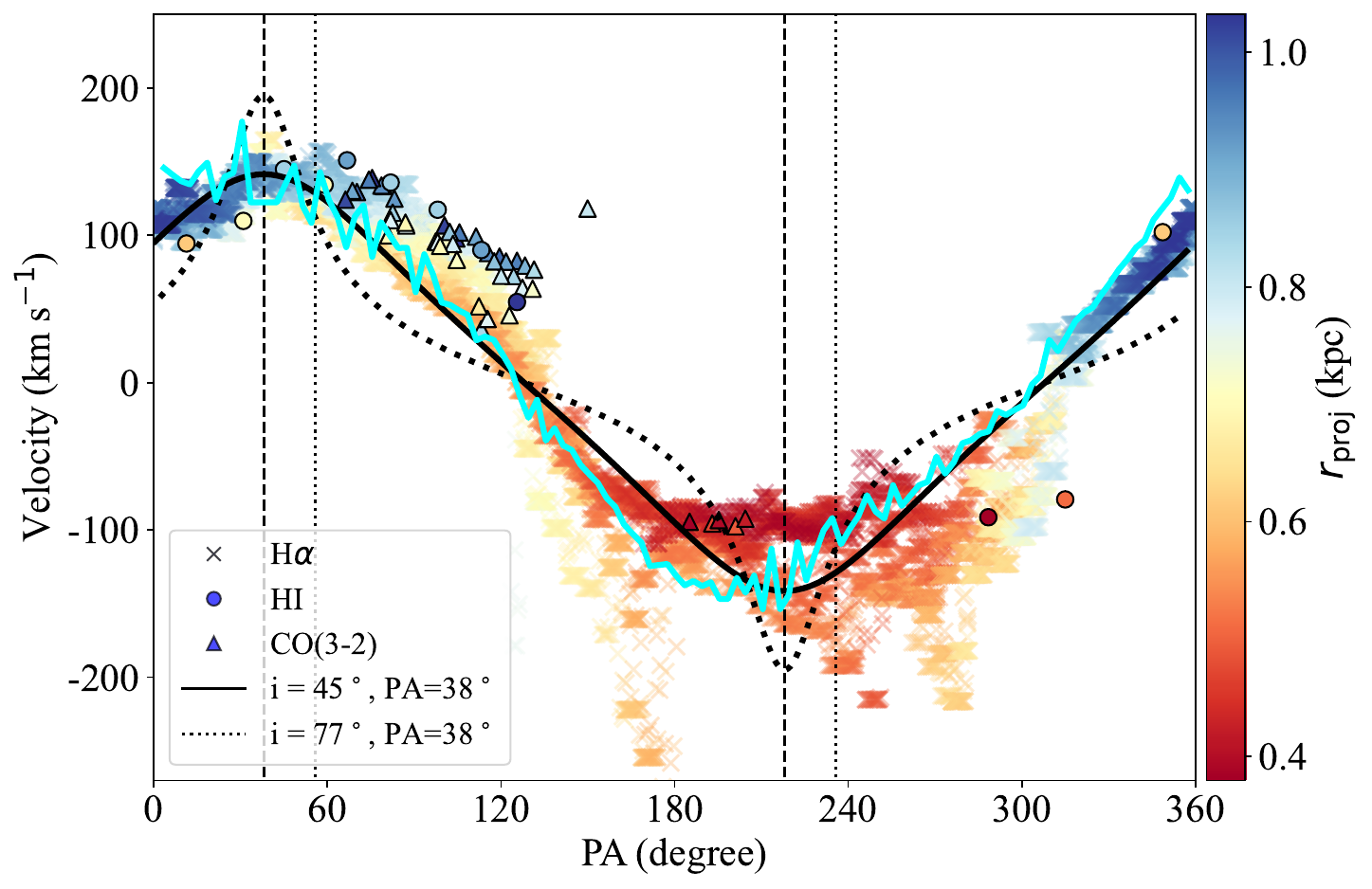}
\caption{Line-of-sight velocity distribution extracted from the quasi-nuclear ring region. Different gas phase images were regridded according to their resolutions. Pixel-by-pixel velocity of H$\alpha$, HI, and CO(3-2) are represented with crosses, circles, and triangles, color-coded with projected distance from the M31 center.
The velocities were corrected for the systemic velocity of M31 ($-300\rm~km~s^{-1})$.
The dotted black line represents a circular rotation pattern with a rotation velocity of 200 km s$^{-1}$ %(in rough consistency with the circular velocity of the stellar disk in Fig. 28 in Blana Diaz 2018?) 
and the same PA (38$^\circ$) and inclination angle (77$^\circ$) as the outer disk. The solid black line represents a similar rotation curve but with a lower inclination angle of 45$^\circ$ \citep{2011A&A...536A..52M}. The cyan curve is the average velocity distribution extracted from the region shown in the lower right panels in Figure \ref{fig:proj}, which is derived from the grid-based hydrodynamical simulation (see Section \ref{sec:pluto} for details). The major axis of the disk is shown with a pair of vertical dashed lines, while that of the bar is shown with a pair of dotted lines. 
\label{fig:vel_dist}}
\end{figure}

To quantify the motion of the quasi-nuclear ring, we derive the azimuthal distribution of the line-of-sight velocity along the ring, as shown in Figure \ref{fig:vel_dist}. 
The definition of the position angle is the same as in Figure~\ref{fig:mass_dist}, and similarly, the velocity profile of HI and CO is under-representative in the SW segment (PA $\sim$ 150$^\circ$--300$^\circ$) of the ring, where H$\alpha$ is most significant.
Nevertheless, the velocity profiles of the three gas phases are in rough agreement with each other, and when combined together, exhibit a roughly sinusoidal shape, characteristic of a planar rotation curve. 
However, the velocity profile of the ring does not precisely match the velocity curve of a rotating circular ring with an inclination angle of 77$^\circ$ (i.e, having the same inclination angle and major axis as the stellar disk of M31) at a speed of 200 km s$^{-1}$, as shown by the black dotted line in Figure \ref{fig:vel_dist}. At large inclination angles, the velocity curve of a rotating circular ring should exhibit narrow peaks near the major axis, whereas all three gas phases show a flat-topped velocity profile. For comparison, we also plot in Figure~\ref{fig:vel_dist} the velocity curve (solid black line) expected for a rotating circular ring viewed at an angle of 45$^\circ$ at a speed of 200 km s$^{-1}$, which more closely resembles the observed velocity profiles, especially in PA $\sim$ 320$^\circ$--60$^\circ$.
Perhaps coincidentally, this speed is similar to the predicted circular rotation speed at a galactocentric radius of 1 kpc based on the dynamical modeling of M31 by \citet{2018MNRAS.481.3210B}. 
A progressively lower inclination angle toward smaller galactocentric radii (albeit not as close to the galactic center as achieved here) was also reported for the HI disk of M31 by \citet{2009ApJ...705.1395C}.
Between PA $\sim$ 60$^\circ$--130$^\circ$, the velocity of the neutral gas is slightly but systematically higher than that of the ionized gas at a given PA, where the latter shows a notable scatter. A closer examination of the velocity field in this region suggests a moderate radial gradient, in the sense that gas lying closer to the M31 center has on average a somewhat lower velocity.

At PA $\sim$ 150$^\circ$--280$^\circ$, there is limited information about the neutral gas in the ring. 
%with noticeable discontinuities evident in the HI PV diagram (see Appendix \ref{sec:HI_spec}). 
On the other hand, the velocity distribution of the ionized gas exhibits a plateau over this PA range, deviating substantially from the expected velocity curve of a coherently rotating ring. 
This discrepancy reinforces the disparity between the SW segment and the NE arc of the ring. 
Notably, the velocity distribution of the ionized gas in this portion is complex and shows multiple distinct features, in particular, peaks around 170, 250, and 280 degrees. These features suggest intricate gas streaming motions superposed on the bulk rotational pattern. 

\begin{figure}[ht!]
\includegraphics[width=\textwidth]{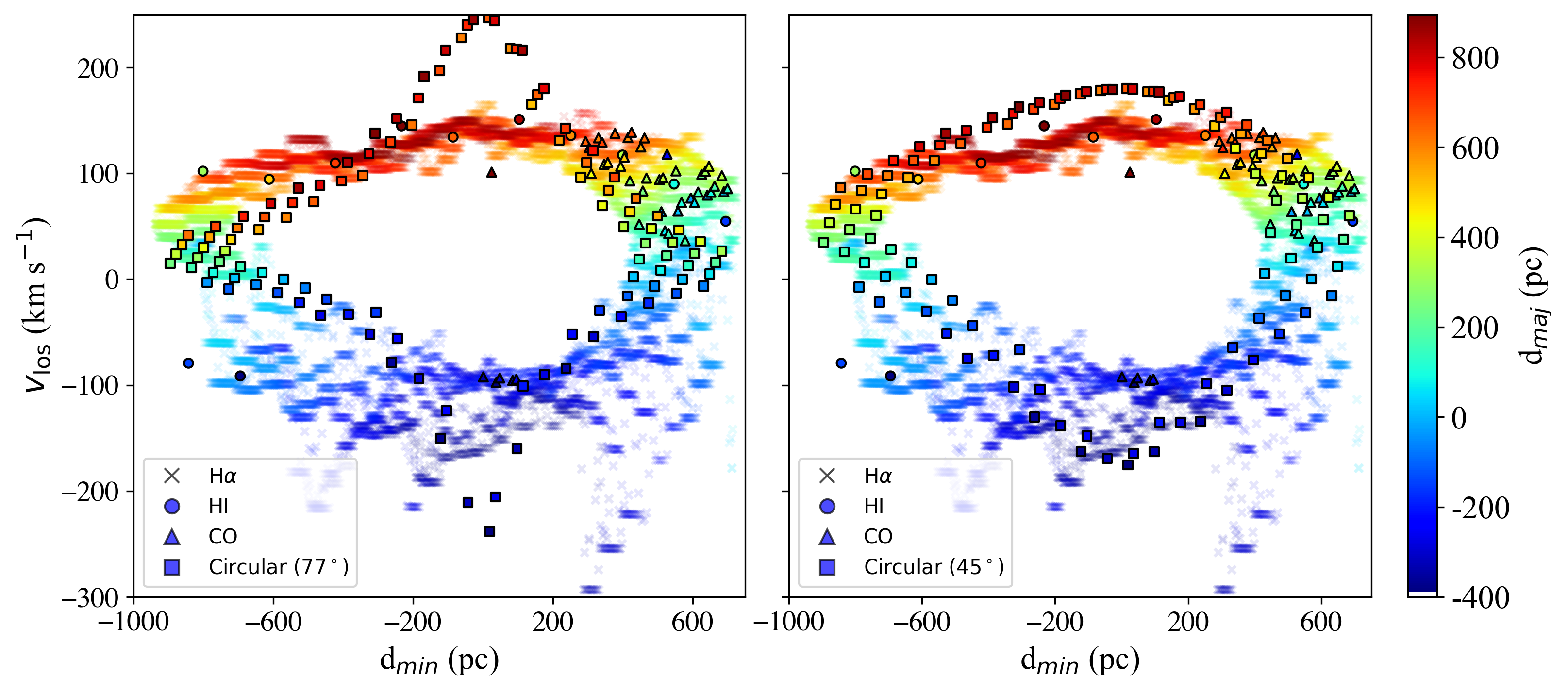}
\caption{Line-of-sight velocity distribution of the ring along the minor axis of the galaxy, with a PA of 128$^\circ$. The velocity fields of H$\alpha$, HI, and CO are represented by crosses, circles, and triangles, respectively. The points are color-coded based on their distance from the center along the major axis with a PA of 38$^\circ$. The square points, displaying smooth and continuous distributions, correspond to pure circular motions observed at a PA of 38$^\circ$ and different inclination angles, 77$^\circ$ ($Left$) and 45$^\circ$ ($Right$), respectively. These velocities are derived from the rotation curve of an analytical model that reproduces the observed properties of M31 well \citep{2006MNRAS.366..996G}, which we also adopted to mimic the M31 potential in Section \ref{sec:pluto}. 
\label{fig:vel_min}}
\end{figure}

To further characterize the velocity field of this quasi-ring, we derived the line-of-sight velocity distribution of the ring along the minor axis, as shown in Figure \ref{fig:vel_min}. The distance is the projected distance along the minor axis away from the galactic center. The points are color-coded with the distance along the major axis, and thus the reddish color represents the NE half, while the blueish color represents the SW half. The distribution of the multi-phase gas exhibits similar characteristics and generally agrees with each other. While the velocity of the NE arc is approximately symmetric about the center, the SW half is highly asymmetric. It exhibits several peaks, indicating intricate gas streams distinct from the NE arc. This is another evidence that this quasi-ring structure is not a coherent ring. The velocity distribution of a pure circular motion based on the galaxy's rotation curve \citep{2006MNRAS.366..996G} is also shown in Figure \ref{fig:vel_min} for a better comparison. The distribution with a PA of 38$^\circ$ and an inclination angle of 77$^\circ$ shows a star-like shape, while that with the same PA but a lower inclination angle of 45$^\circ$ results in a rounder one. The latter is in rough consistency with the observed distribution of the NE arc, suggesting that the NE arc should have a lower inclination angle than the outer disk, consistent with previous conclusions.

{\vskip0.5cm}
To summarize, the multi-wavelength observations presented above have revealed several important aspects of the quasi-nuclear ring: (1) the molecular phase likely dominates the overall mass budget, (2) a substantial variation in the surface mass density across the ring is seen in the molecular, atomic and ionized phases, (3) the NE arc and SW segment of the ring is primarily manifested by the neutral phase and ionized phase, respectively, and (4) the velocity field of the NE arc of the ring is compatible with a circular motion %superposition of circular and radial motions 
viewed at a substantially more face-on angle than the outer disk, whereas the velocity field of the SW segment of the ring, essentially traced by the ionized gas, appears more complex than such a simple bulk motion.    
These newly determined properties of the quasi-nuclear ring, in particular (2)--(4), challenge the case of this ring-like feature being a coherent entity. In the next section, we attempt to provide a plausible physical picture of this nuclear structure, demonstrating that it is more likely a set of spirals rather than a coherent ring.

\section{Simulations}
\label{sec:sim}

In this section, we performed a series of custom hydrodynamical simulations to investigate the evolution of the gas disk of M31. These simulations focus on the effects of the M31-M32 collision and the presence of a bar, two leading scenarios proposed to explain the observed nuclear gas structures (Section \ref{sec:discussion}). Although bar-induced spirals have been investigated in detail in previous theoretical studies \citep[e.g.][]{1956StoAn..19....2L, 2012ApJ...758...14K, 2012ApJ...747...60K}, a custom hydrodynamical simulations of the M31-M32 collision incorporating a bar is still missing. N-body/SPH simulations are performed to investigate the collision in a self-consistent manner using a self-evolving gravitational potential. In contrast, pure hydrodynamical simulations are used to analyze the gas behavior in a more realistic potential with parameters constrained with observations \citep{2006MNRAS.366..996G}. We stress that the main aim of these simulations is to illustrate the qualitative behavior of gas components in the central region of M31, based on its well-established global properties and probable past interaction with M32. 
A more thorough exploration of the many relevant parameters is beyond the scope of the present study.

\subsection{N-body/SPH simulations}\label{sec:SPH}
%To explain the observed gas structures in the nuclear region, 
We utilized the $N$-body/SPH simulation tool, GADGET-4 \citep{2021MNRAS.506.2871S} to investigate the M31-M32 collision under different impact parameters in a self-consistent way. 
M31 was modeled with a Navarro-Frenk-White (NFW) dark matter halo \citep{1996ApJ...462..563N, 1997ApJ...490..493N} plus an exponential stellar disk and a Hernquist stellar bulge \citep{1990ApJ...356..359H}, with a total mass of 1.6 $\times 10^{12}\rm~M_\odot$  \citep{2002ApJ...573..597K}. The M31 disk had a total baryonic mass (including the stellar and gas components) of 8 $\times 10^{10}\rm~M_\odot$, with initial scale heights of approximately 1 kpc and 0.1 kpc for the stellar and gas components, respectively. The scale lengths for the M31 disk and bulge are 5.5 and 1.8 kpc, respectively.  
M32 was modeled with an NFW dark matter halo and a Hernquist bulge, with a total mass of 3$\times 10^{10}\rm~M_\odot$ and a scale length of 0.3 kpc. No gas component was assigned to M32. Regarding resolution, the mass of individual dark matter particles was 2$\times 10^5\rm~M_\odot$, while the mass of individual stellar or gas particles was 1$\times 10^4\rm~M_\odot$. This results in 17{,}935{,}000 particles in total for all the models. These simulations utilized the default built-in recipes of GADGET-4, which include cooling, star formation, and stellar feedback (but no AGN feedback).

After evolving M31 and M32 in isolation for 2 Gyr to achieve dynamical relaxation, we put them into the same frame and performed five sets of simulations, each with a certain combination of the initial position and velocity of M32, corresponding to (1) a head-on collision with a relative velocity of $\sim$400 km s$^{-1}$ (slightly higher due to the free-fall of M32); 
(2) a head-on collision with a lower velocity of $\sim$200 km s$^{-1}$; (3) an oblique central collision with an incident angle of $\sim 30^\circ$ and a relative velocity of $\sim 300\rm~km~s^{-1}$; (4) an off-centered collision with an in-disk impact distance of 2 kpc, an incident angle of $\sim 30^\circ$ and a relative velocity of $\sim 300\rm~km~s^{-1}$. In simulation set (5), M31 was evolved in isolation for 5 Gyr, and a bar was generated spontaneously, with a pattern speed of $\sim 20\rm ~ km~s^{-1}~kpc^{-1}$ and a length of $\sim 5$ kpc. Then M32 was introduced into the system with initial conditions the same as in the simulation set (4). 
%The initial positions of M31 were all at (0, 0, 0) kpc, with initial velocity of (0, 0, 0) km s$^{-1}$, 
The parameters of the five simulations are summarized in Table \ref{tab:model}. In all simulations, M31 was initially at rest at the origin, and the initial disk normal is at (0, 0, 1), that is, the $+z$ direction. During impact, only a small portion of the particles belonging to M32 are stripped and accreted into M31, which has little effect on the results.

\begin{table}[h!]
    \centering
    \begin{tabular}{cccccc}
    \hline
        Model & M32 Initial position & M32 Initial velocity & In-disk impact position & Incident angle &  Bar \\
        &(kpc)&(km s$^{-1}$) & (kpc) & ($^\circ$) &\\
         \hline
       (1)  & (0, 0, -25) & (0, 0, 400) & 0 & 0 & No \\
       (2)  & (0, 0, -25) & (0, 0, 200) & 0 & 0 & No \\
       (3)  & (14, 0, -25) & (-150, 0, 260) & 0 & 30 & No \\
       (4)  & (16, 0, -25) & (-150, 0, 260) & 2 & 30 & No \\
       (5)  & (16, 0, -25) & (-150, 0, 260) & 2 & 30 & Yes \\
    \hline       
    \end{tabular}
    \caption{Hydrodynamical model parameters}
    \label{tab:model}
\end{table}

\begin{figure*}[!hbtp]
	\centering
	\includegraphics[width=\textwidth]{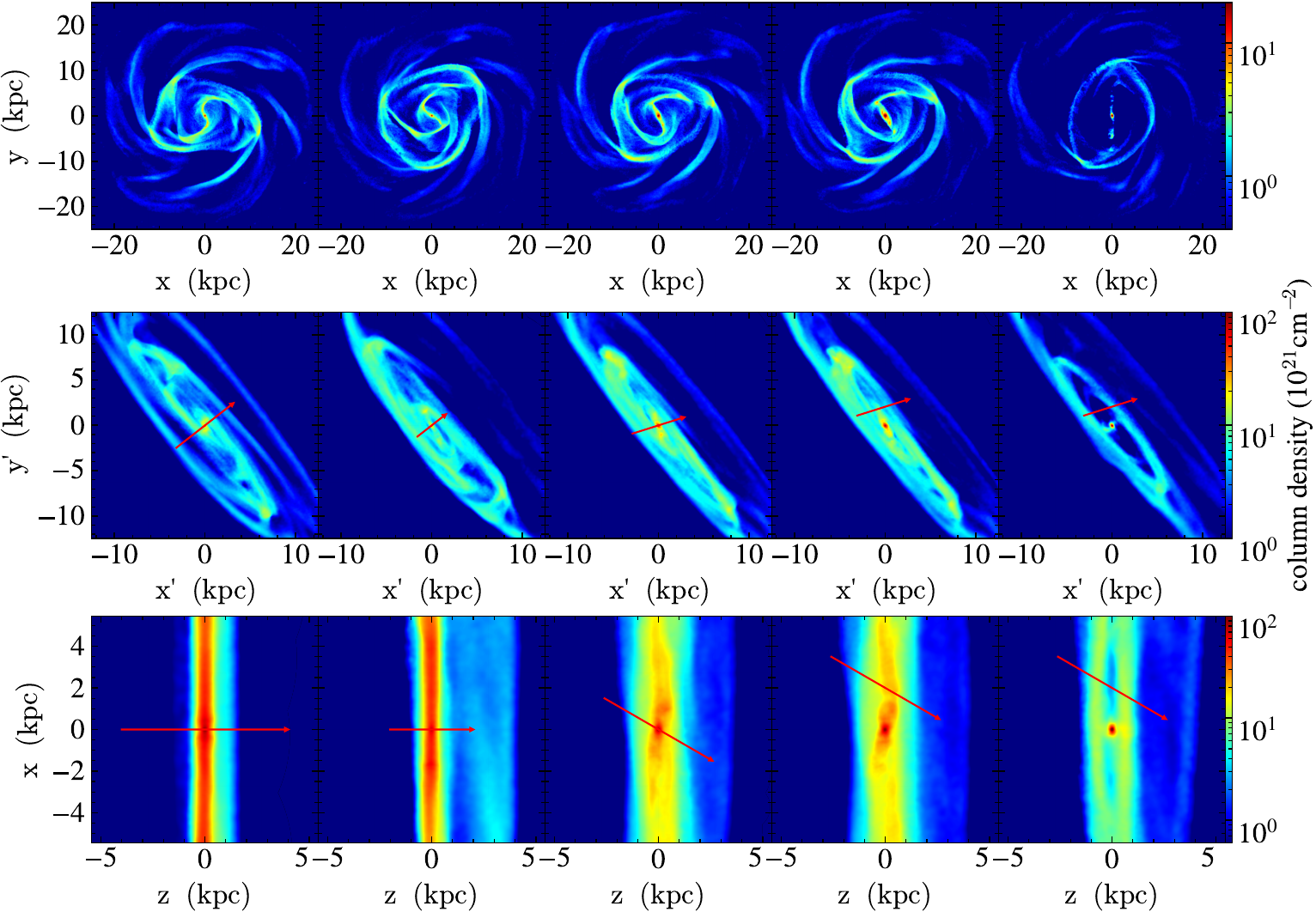}
	\caption{Snapshots of the whole gas disk of M31 $\sim$400 Myr after the mid-plane passage of M32 in the N-body/SPH simulations. The panels from left to right correspond to simulations (1)--(5). The top and middle panels show the face-on and projected views, respectively. The projection adopted an inclination angle of $77^\circ$ and a position angle of $38^\circ$, i.e., in the same way as M31 is seen by us. The bottom row shows the edge-on view of the central 5 kpc. The collisional direction of M32 is marked with red arrows, where the length is scaled with the relative velocity. 
\label{fig:gadget4}}
\end{figure*}

All simulations produced sustained outward-propagating collisional rings in the disk plane. %(Figure~\ref{fig:rings}). 
Representative snapshots of the global view of the simulated gas component of M31 are shown in Figure~\ref{fig:gadget4}.
In the direction perpendicular to the disk plane, simulation (1) shows no pronounced oscillations in the central region. The oscillations in the simulation (2) were more evident due to the lower relative velocity and the longer duration of the interaction. In the remaining three simulations, a significant torque occurred due to the oblique incidence, resulting in the most pronounced vertical perturbation of the gas disk. In particular, significant twists in the central region are produced, as can be seen in the zoom-in, edge-on view in Figure \ref{fig:gadget4}. Compared to model (3), model (4) demonstrates that an off-centered oblique collision can generate significant asymmetry as observed. Furthermore, only model (5) successfully reproduces the prominent outer ring at 10 kpc as observed, as the bar potential clears the gas within this radius and reinforces the ring through resonance. %Zoomed-in views of the simulations are presented in Figures \ref{fig:SPH_zoomin} and \ref{fig:SPH_bar}.

\subsection{Grid-based Hydrodynamical Simulation}\label{sec:pluto}

Based on the $N$-body/SPH simulations, only an oblique collision can produce a tilted central region, as suggested by the observed ring morphology and velocity field. Furthermore, the presence of a bar is essential for forming a distinct 10 kpc ring and effectively sweeping out the gas within this radius, as has been well established in observations \citep[e.g.,][]{2009ApJ...695..937B}. However, the bar potential in model (5) is self-generated and may not accurately represent the actual bar properties of M31. Additionally, the limited resolution of the $N$-body simulation prevents the clear identification of nuclear substructures.
To resolve the joint effect of bar-driven inflows and the collision on the nuclear substructures, we performed grid-based hydrodynamical simulations with the publicly available code {\sc pluto} \citep{2007ApJS..170..228M}. This also enables us to adopt a custom potential for the M31 galaxy described by an NFW dark matter halo, an exponential disk, a Hernquist bulge, and a central black hole point mass, which reproduces the observed properties of the galaxy well \citep{2006MNRAS.366..996G}. 
We adopt a three-dimensional static Cartesian grid. In the $x$--$y$ plane, i.e., the galactic disk plane, 512 cells are uniformly spaced to cover $\rm [-12~kpc,~12~kpc]$ along each axis. Along the $z$-axis, 15 cells are uniformly spaced to cover $\rm [-0.28~kpc,0.28~kpc]$, while the ranges of $\rm [-12~kpc, -0.28~kpc]$ and $\rm [0.28~kpc, -12~kpc]$ are logarithmically spaced with 30 cells each.
These settings allow the gas structure and kinematics in the central kpc to be resolved well while covering a sufficiently large outer region. 
We assume an isothermal equation of state with a sound speed $c_{\rm s}=10~\rm km~s^{-1}$, and we neglect effects such as star formation and feedback.
Our simulations consist of two stages: (i) the bar potential implementation and development of bar-driven gas structures, called the ``bar stage"; (ii) the collisional passage of M32 and subsequent evolution of the M31 gas, called the ``collision stage". 

For the ``bar stage", the bar is implemented with a quadrupole potential (see more details in \citealp{2024MNRAS.528.5742S}) with pattern speed $\Omega_{\rm p}=40~\rm km~s^{-1}~kpc^{-1}$ and bar strength $Q_{\rm b}=0.1$ \citep{2018MNRAS.481.3210B}.
The ``bar stage" simulation is run for 600 Myr to achieve a quasi-steady state. %as illustrated in Figure~\ref{fig:bar}.
At this point, the bar has driven gas inflows and the gas stabilizes around the classical $x_2$ orbits.  

\begin{figure*}[!hbtp]
	\centering
	\includegraphics[width=1.0\textwidth]{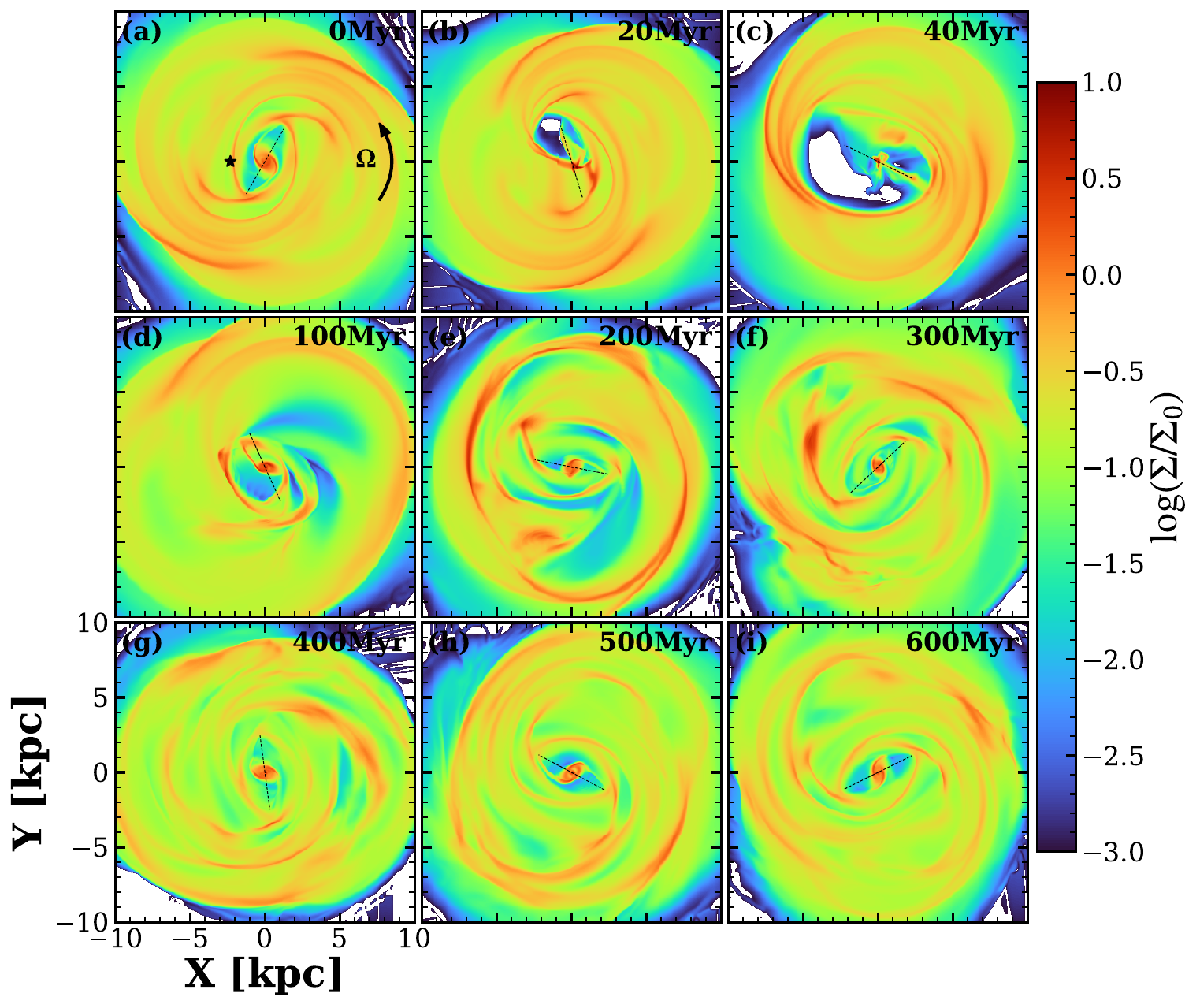}
	\caption{Surface mass density distribution of the ``collision stage" 3D simulations. The time of the snapshots presented in each panel is relative to the pericenter passage of M32, which has a pericenter distance of $R_{\rm p}=2\rm ~kpc$ and an incident angle of $30^\circ$. The position of the collision is denoted by the star in panel (a). 
 %The color bar denotes the mass-weighted density in logarithm, projected on the disk plane. 
   The major axis of the bar is illustrated by a black dashed line, which has a counterclockwise rotation. 
 \label{fig:collision}}	
\end{figure*}

\begin{figure*}[!hbtp]
	\centering
	\includegraphics[width=1.0\textwidth]{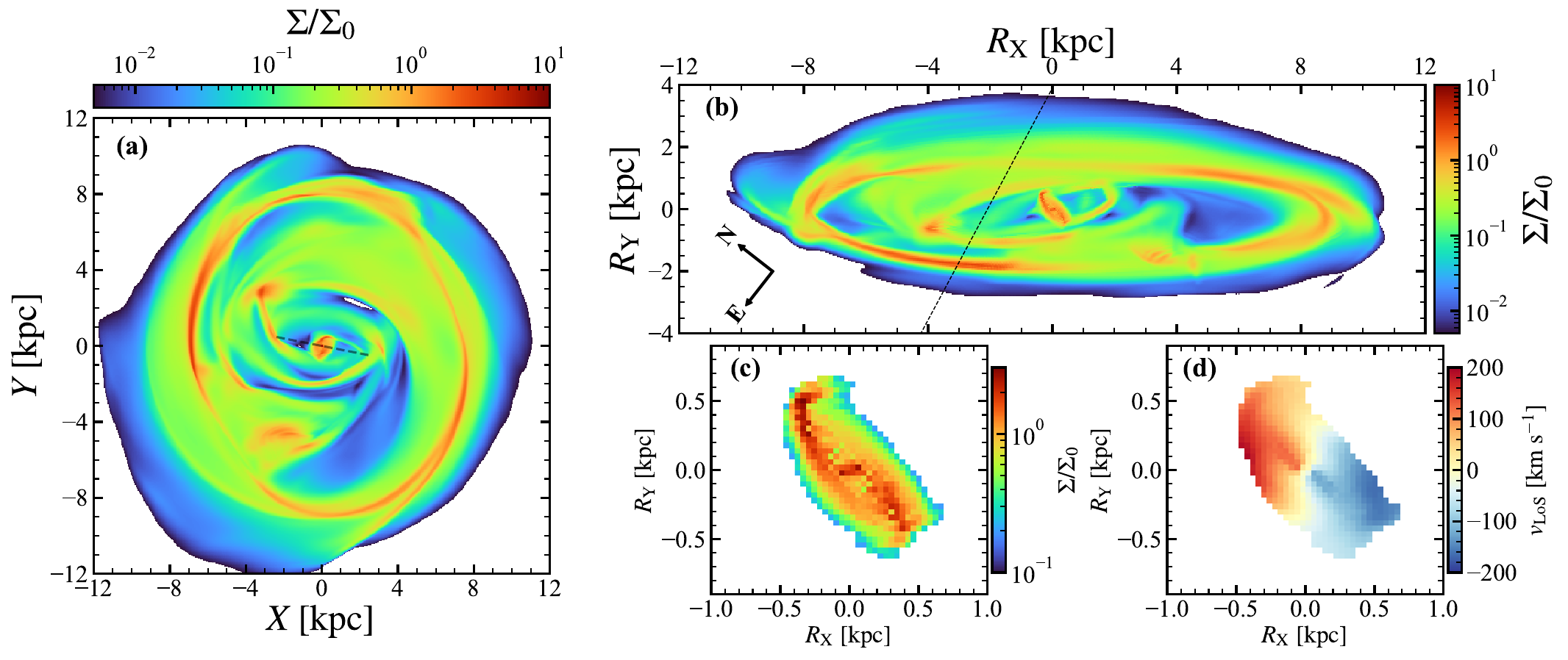}
	\caption{Snapshot of the ``collision stage" 3D simulations at $\sim 200$ Myr after the pericenter passage of M32. (a) Surface mass density distribution of the face-on view. The major axis of the bar is illustrated by a dashed black line. (b) Projected surface mass density distribution with an inclination angle of $77^\circ$ and a bar angle offset from the disk's major axis by 17.7$^\circ$ \citep{2017MNRAS.466.4279B}. The trajectory of M32 is marked as the dashed black line. (c) The zoomed-in view of the surface mass density distribution and (d) the velocity field of the nuclear gas structure.}
    \label{fig:proj}
\end{figure*}

For the ``collision stage", an off-centered and inclined collision is assumed. The intruder has an impact parameter of 2 kpc and a constant velocity of $300~\rm km~s^{-1}$ with an impact angle of $30^\circ$ relative to the symmetry axis of the galactic disk, identical to the setup of model (5) in Section \ref{sec:SPH}. The potential of the intruder is mimicked using a Plummer potential with a mass of $2\times10^{10}~\rm M_\odot$ \citep{2014ApJ...788L..38D} and a radius of $0.1~\rm kpc$ 
\citep{2002ApJ...568L..13G}, similar to the observed properties of M32. It should be noted that since the perturbation induced by M32 is small, the choice of the specific potential does not have a significant influence on the results.
%To mimic the potential of the intruder galaxy M32, we use a Plummer potential with a mass of $2\times10^{10}~\rm M_\odot$ \citep{2014ApJ...788L..38D} and Plummer radius of $0.1~\rm kpc$ \citep{2002ApJ...568L..13G}. 
We note that, for simplification, the intruder's mass and structure do not evolve in the simulation. The ``collision stage" is run for 800 Myr when the perturbation in the central region is nearly fully relaxed.
During this stage, collisional rings and tilted spirals emerge due to perturbation from the off-centered encounter. As shown in Figure \ref{fig:collision}, the first structure developed around 20 Myr after the pericenter passage is an incomplete expanding ring. The expanding velocity is about $80~\rm km~s^{-1}$, thus this ring would leave the box within 150 Myr. After 200 Myr, the subsequent collisional rings are more complete and circular with a lower expanding velocity of $15-30~\rm km~s^{-1}$. In the inner 3 kpc, the bar-induced structures like spirals and trailing waves are significantly disrupted by the collision. It is worth noting that in contrast to the observed deficiency of gas in the central region, gas is accumulated within the nuclear spirals in our simulation. We attribute this disparity to the absence of feedback processes in our simulation, which would otherwise deplete or expel gas. As the intruder runs away, the amplitude of the perturbation damps, and the original bar-driven structure gradually resumes in the central region. During the intermediate phases (about $200-600~\rm Myr$), a pair of spirals are present, tilted due to the torque of the encounter, such that they have a lower inclination than the galactic disk and together mimic a ring, as demonstrated in Figure \ref{fig:proj}.

%\newpage

\section{Possible origins of the nuclear gas structure}\label{sec:discussion}

The origin of gas structures in the central kpc region of M31, including the quasi-nuclear ring and the nuclear spiral, has long been investigated and discussed in the literature \citep{1994ApJ...426L..31S, 2011A&A...536A..52M, 2022ApJ...933..233F, 2023ApJ...953..109A}. Both analytical methods and numerical simulations were applied, focusing on two main scenarios: 
%several hypotheses, including 
gas inflows driven by a triaxial bulge or barred potential \citep{1994ApJ...426L..31S}, or a collisional ring produced by a recent close-in passage of a satellite, most likely M32 \citep{2006Natur.443..832B, 2011A&A...536A..52M}. 
%a tilted ring plus an inclined disk \citep{2011A&A...536A..52M}, 
The formation of the nuclear spiral was recently modeled as gas streams from a pre-existing nuclear ring \citep{2023ApJ...953..109A}.
However, these studies have been hindered by a lack of comprehensive insight into gas distribution and kinematics throughout the entire nuclear region. 
%due part to the scarcity of gas. 
Leveraging our new CO observations as well as the sensitive HI and H$\alpha$ maps presented above, we can now construct a relatively holistic perspective of the morphology and kinematics of multiphase gas across the quasi-nuclear ring. 
The multi-wavelength data suggest that the gas in this area can be best described as an off-centered, lopsided two-armed spiral structure tilted by $\sim 30^\circ$ with respect to the outer disk. In the following, assisted by custom numerical simulations, we aim to unravel the possible origin(s) of this gas structure with a semi-quantitative comparison with the observations.

\subsection{Bar-driven inflow scenario}
\label{subsec:bar}
That a bar is probably present in M31 \citep{1977ApJ...213..368S, 2006MNRAS.370.1499A, 2007ApJ...658L..91B} provides a plausible explanation for the nuclear structure. The bar potential can create close orbits, known as the $x_1$ and $x_2$ orbits \citep{1977A&A....61..477C}, and produce dust lanes as gas transit from outer $x_1$ orbits to inner $x_2$ orbits, and/or nuclear rings, as gas concentrates around the $x_2$ orbits \citep{2012ApJ...747...60K}. Evidence for a bar in M31 is present in observations of both stellar and gaseous components. In recent made-to-measure $N$-body simulations \citep{2017MNRAS.466.4279B, 2018MNRAS.481.3210B}, a stellar bar with a half length of 4 kpc and a pattern speed of $\Omega\rm_p \approx 38 ~km~s^{-1}~kpc^{-1}$ is required to simultaneously explain the 3.6 $\mu$m photometry \citep{2006ApJ...650L..45B}, stellar kinematics \citep{2018A&A...611A..38O} and the HI rotation curve \citep{2010A&A...511A..89C}. 
\citet{1994ApJ...426L..31S} found that the velocity field then observed of circumnuclear neutral and ionized gas can be explained by a weak bar potential, in particular, the ``face-on spiral'' of ionized gas (i.e. the SW segment of the nuclear structure discussed in Section~\ref{sec: results}) can be understood as the transition region as gas moves from $x_1$ orbits to $x_2$ orbits, which is often associated with shocks. 
\citet{1994ApJ...426L..31S} argued that this transiting feature (also known as dust lanes) could appear round if the $x_1$ orbits are elongated close to the line-of-sight. However, the velocity data available to \citet{1994ApJ...426L..31S} were much more limited than presented here, and their inferred bar parameters (orientation and pattern speed) differ substantially from modern values.
More recently, sharp [O\,{\sc iii}] velocity discontinuities in gas kinematics were revealed by VIRUS-W IFU observations \citep{2018A&A...611A..38O}, which are interpreted as dust lane shocks associated with the $x_1$ orbits \citep{2022ApJ...933..233F}. However, these velocity discontinuities lie significantly outside the nuclear structure, and a slower bar pattern speed ($\sim 16-20 ~\rm km~s^{-1}~kpc^{-1}$) is preferred in the dynamical gas model, although the result depends on the uncertain inclination angle of the central region \citep{2022ApJ...933..233F}. 

We consider two possible cases of ring-like structures related to the bar. In the first case, the inflowing gas is manifested by a pair of elongated ``arms'' or dust lanes across the family of $x_1$ orbits, which look round and resemble a ring-like feature after projection, as originally suggested by \citet{1994ApJ...426L..31S}. 
One of the two main problems with this interpretation is that the overall orientation of the two segments of the quasi-nuclear ring appears to be roughly perpendicular, rather than parallel, to the inferred major axis of the bar (Figure~\ref{fig:fov} and \ref{fig:flux}), which is difficult to reconcile with the $x_1$ orbits that align with the bar. The other problem lies in that the quasi-nuclear ring is highly asymmetric in terms of the surface mass density and substantially off-centered, as idealized simulations of a bar potential generally predict gas streams symmetric about the center, in accordance with the point symmetric potential. 
Even in more realistic hydrodynamical simulations accounting for feedback processes, the resultant nuclear spirals are still point symmetric \citep{2023arXiv231104796F}.
Although recent work suggested that asymmetric gas flows can develop spontaneously due to hydrodynamical instabilities \citep{2018MNRAS.475.2383S}, which may potentially explain the asymmetric surface gas density observed both in the CMZ and in M31's quasi-nuclear ring, it remains unclear how such a mechanism could generate an off-center pattern as observed in M31.

The second case involves a bar-induced nuclear ring lying between the inner inner and outer inner Lindblad resonances \citep[IILR and OILR,][]{1956StoAn..19....2L}, co-spatial with the family of $x_2$ orbits (\citealp{2012ApJ...758...14K}). For M31, the IILR and OILR in the bar potential model are located at a galactocentric radius of 1 kpc and 1.8 kpc, respectively (Figure \ref{fig:fov}, \citealp{2017MNRAS.466.4279B, 2018MNRAS.481.3210B}), which are compatible with the location of the observed quasi-nuclear ring. 
However, that the ring is off-centered remains a severe challenge to this scenario. Moreover, canonical $x_2$ orbits aligned with the disk plane cannot reproduce the observed flat-topped velocity curve of the ring, which suggests a tilt of approximately $30^\circ$ relative to the outer disk (Section~\ref{subsec:kinematics}).
Although simulations have found that bars can induce gas inflows with tilts up to 
%of 1$^\circ$ --
5$^\circ$ in the CMZ \citep{2020MNRAS.499.4455T}, this is insufficient to account for the substantial tilt inferred for the nuclear ring of M31.
Furthermore, any strong asymmetry in the gas phases across the ring should be erased within the dynamical timescale (rotation period) of a resonant ring, which is only 30 Myr, assuming a circular speed of 200~km~s$^{-1}$ (Section~\ref{subsec:kinematics}).

Therefore, we disfavor the canonical bar-driven inflows as the sole origin of the nuclear structure.

\subsection{Collisional ring scenario}
\label{subsec:ring}
Ring-like features present in galaxies may also be caused by a nearly head-on collision with a satellite galaxy \citep{1976ApJ...209..382L}, the so-called collisional rings \citep{1996FCPh...16..111A}, the manifestation of sequential, nearly point-symmetric density waves driven into a disk in response to the radial oscillation of the perturbed gravitational field. In the specific case of M31, 
%it is believed that355 the satellite galaxy responsible for this is M32
it is suggested that a recent passage of M32 through the central region of M31 has produced a series of two collisional rings,  
%The fact that M31's two main rings, 
located at 0.7 kpc and 10 kpc, respectively, and off-centered by $0.5-1$ kpc \citep{2006Natur.443..832B}. No bar is required in this scenario.
%suggests disturbance from satellite galaxies. 
Several investigations have endeavored to simulate the impact of the M32 collision and determined that such a collision could generate rings with characteristics resembling those observed, such as the ring centers being displaced from the galactic center, a significant tilt of the inner ring out of the disk mid-plane, and the presence of a void within the 10-kpc ring \citep{2006Natur.443..832B, 2006ApJ...638L..87G, 2014ApJ...788L..38D}.

We then utilized our custom SPH simulations to explore the effect of a pure collision with M32 on the gas morphology in M31. %in particular, how collisional rings form and evolve and to what extent the inner ring can be tilted.
The simulations include realistic contents of dark matter and stars in both M31 and M32 and a gas disk in M31 (no gas for M32). 
After having M31 first evolve in isolation to achieve dynamical relaxation, spiral arms are generated spontaneously during the secular evolution phase, but no bars or rings are initially implemented. M32 is then introduced as a compact intruder from a remote distance and subsequently collides with M31 at various initial velocities, impact positions on the disk, and incident angles as discussed in detail in Section \ref{sec:SPH}. 
All our simulations result in a sequence of two to three expanding ring/spiral features between 100--500 Myr after the collision, as discernible in Figure \ref{fig:collision}.

A collision could cause a significant tilt in the nuclear gas structure, which would be consistent with the low inclination angle of the quasi-nuclear ring inferred from observation. In accordance with \citet{2006Natur.443..832B}, our simulations suggest that a low relative velocity %, an impact position close to the galactic center, 
and a moderately oblique incident angle is required to generate the large tilt suggested by the gas morphology in the central region (Figure~\ref{fig:gadget4}). 
In particular, a relative velocity on the order of $200-300$ km s$^{-1}$ is necessary to produce significant oscillations in the disk due to the longer interaction time with M32 \citep{2012MNRAS.425.2255F}. A substantial oblique incident angle (e.g. 30$^\circ$) is the key to generate a significant tilt of the nuclear gas structure with respect to the outer disk. On the other hand, a larger impact position on the disk (e.g. 2 kpc) can still cause a substantial tilt in the central region. 
%which is compatible with the observed lower inclination angle of the nuclear ring. 
Additionally, an off-centered oblique collision can induce further asymmetry in the system, providing a plausible explanation for the off-centered rings in M31 \citep{2006Natur.443..832B, 2014ApJ...788L..38D}.

The collisional ring scenario thus seems to offer some advantages in explaining the observed characteristics.
However, this scenario is not without challenges. After the collision, it is evident that a pure collision cannot sweep up gas inside the central region effectively and generate a distinct outer ring. Although collision can cause a significant tilt in the central region as observed, gas accumulates inside the 10 kpc ring, and none of the pure collision models is reminiscent of the two-ring structure (model 1-4 in Section \ref{sec:SPH}). 
In addition, our simulations predict an expansion velocity of the first collisional ring as large as 50--60 km s$^{-1}$, and 30--40 km s$^{-1}$ even for the second or third collisional rings. Given such a rapid expansion, the collisional ring exhibits significant turbulence and takes time to claim a circular shape, rendering it hard to be observed as a ring at small radii in our simulations. Moreover, the pure collisional ring model makes it difficult to reproduce the sharp [O\,{\sc iii}] velocity jumps identified in the ionized gas at several kpc from the galactic center \citep{2018A&A...611A..38O}. The positions of the velocity discontinuity follow a spiral pattern rather than a circular pattern \citep{2022ApJ...933..233F}, which can be more naturally explained with bar-induced shocks but is hardly compatible with collisional rings. The presence of feather-like substructures in the NE arc (Figure \ref{fig:optical}) also supports the notion of a bar-induced curved shock (Section \ref{subsec:surfacemap}), whereas a collisional ring would typically exhibit a smoother structure with significant expansion.

\subsection{Hybrid scenario}
\label{subsec:hybrid}
Regarding the aforementioned difficulties in the bar-only and collision-only scenarios, we examine a hybrid scenario that involves a barred potential disturbed by a satellite intruder, namely M32. After letting M31 evolve in isolation for 5 Gyr in the $N$-body/SPH simulation, a stellar bar with a pattern speed of $\sim 20\rm~km~s^{-1}~kpc^{-1}$ emerges spontaneously, as shown in the last column of Figure \ref{fig:collision}. We then introduced an off-centered (in-disk impact distance of 2 kpc), oblique ($30^\circ$) collision with M32 to examine the effect of a collision on the bar and subsequent evolution of the gas in the central region. %The parameters are identical to the off-centered oblique collision simulation described in Section \ref{sec:SPH}. %(see details in Appendix \ref{sec:SPH}). 
In this case, the bar is not destroyed after the collision. Rather, it is somewhat disturbed and thickened, and causes gas to accumulate in the central kpc and the 10 kpc ring through resonance. 
This is mainly because the bar effectively sweeps up gas co-spatial with the $x_1$ orbits over the M32 passage, mirroring the observed gas deficiency in the central region. However, there is no guarantee that a bar with properties close to those observed can be generated self-consistently in our $N$-body/SPH simulations.

Therefore, to better quantify the gas behavior in the central kpc, a grid-based hydrodynamic simulation is performed with {\sc pluto} \citep{2007ApJS..170..228M}, in which gas evolves isothermally in a fixed, realistic bar potential \citep[][see Section \ref{sec:pluto} for details]{2006MNRAS.366..996G}. Compared to the $N$-body/SPH simulations in Section~\ref{sec:SPH}, this hydrodynamic simulation can manipulate bar-induced inflows in a well-defined bar potential, with the caveat of neglecting the response of the M31 gravitational potential to the passage of M32.
In this simulation, M31 evolves in isolation for 1.2 Gyr before the collision event. During this period, the bar gives rise to gas features occupying the outer $x_1$ orbits, an inner $x_2$-type ring with a radius of $\sim$0.5 kpc, and spiral-like features in between (Figure \ref{fig:collision}). Subsequently, M32 is introduced into the system as a compact, purely gravitational source, colliding with the M31 disk at an impact distance of 2 kpc from the center at an incident angle of $\sim 30^\circ$, consistent with the $N$-body/SPH simulation. Immediately following the impact, pre-existing structures in the nuclear region, such as the $x_2$-type nuclear ring and nuclear spirals, are disrupted. Over the next few hundred Myr, a sequence of collisional rings forms beyond the central kpc and propagates outward (see Figure \ref{fig:collision}). In agreement with the $N$-body/SPH simulation, even the second and third rings display expansion velocities of 15–30 km s$^{-1}$. As a result, these expanding rings appear ring-like only at larger radii (2–3 kpc), whereas no complete ring morphology is observed at the scale of the quasi-nuclear ring.
Concurrently, the collision induces a significant tilt in the nuclear region, reaching up to 40$^\circ$ within the central 1--2 kpc range.

Despite the fierce perturbation induced by the oblique collision, the disrupted gas gradually re-converges into the central region under the influence of the bar potential after 100 Myr. In our simulation, the converged gas forms a pair of nuclear spirals that resemble the $x_1$ to $x_2$ transition $\sim 100$ Myr after the collision, with a size of $\sim 1$ kpc. This central structure is relatively stable for more than 500 Myr in our simulation. Due to the tilt of the central region, the appearance and ellipticity of the structure vary as it rotates. At certain epochs ($\sim 200$ Myr after the collision in our simulation), a prominent 10 kpc ring plus tightly wounded nuclear spirals emerge, with several disrupted spiral arms in between (Figure \ref{fig:collision}). %the outer part of the spirals appears round after projection, while the projected inner part of the spirals resembles a spoke inside (Figure \ref{fig:proj} and Figure~\ref{fig:sim_NR}).
Remarkably, this configuration mirrors the observed morphology of the M31 disk. In particular, after projection, the tightly wounded nuclear spirals in the simulation appear ring-like and resemble the observed quasi-ring morphology as shown in Figure~\ref{fig:fov}. %within the central kpc of M31, featuring two tightly wound spirals together manifesting a ring-like structure, accompanied by spiral-like filaments inside, as shown in Figure~\ref{fig:fov}. 
The tilt induced by the collision can remain for about 600 Myr, after which the tilted spirals would gradually fall back to the galactic plane and finally stabilize on the $x_2$-orbit under the bar potential.
Furthermore, the simulation exhibits notable variations in surface density throughout this structure, reflecting the asymmetry seen in the azimuthal distribution of gas surface density. It is worth noting that our simulation shows a higher degree of gas concentration in the central few hundred-parsec region. We attribute this disparity to the absence of feedback processes in our simulation, which would otherwise help deplete or expel the gas.

Moreover, the simulation-predicted line-of-sight velocity distribution across the nuclear spirals is also in rough agreement with the observed velocity distribution of the multiphase gas, as illustrated in Figure \ref{fig:vel_dist}. 
The discrepancy between the simulation and the observation in PA $\sim$ 200$^\circ$--300$^\circ$ suggests a more complex structure in this region, likely consisting of a mixed origin rather than originating solely from the bar-induced inflow. One possibility is that the tilt of the central structure causes the outer disk to project onto one side of the two-armed spiral, similar to the geometry described in \citet{2021ApJ...909...81R}. This projection could result in contamination within the position angle range of $\sim$ 200$^\circ$–300$^\circ$, contributing to the apparent off-centered geometry.
%and causing the apparent asymmetric between the two arms.

The hybrid scenario thus offers the advantage of producing a gas structure comprising bar-induced streamers within the central kpc, which is tilted but remains relatively stable for up to several hundred Myr. This stability enhances the likelihood of it being observed in reality. 
Although this scenario also involves uncertainties regarding M32's mass and orbit and the need for a specific viewing angle, it largely replicates the observed quasi-nuclear ring. 
What is not satisfactorily explained, however, is the observed lopsidedness in the predominant gas phase and asymmetric geometry between the NE arc and SW segment of the ring (more precisely, the two tightly-wounded spirals). Figure~\ref{fig:fov} suggests that the two segments are connected to, and thus likely fed by outer gas structures.
At the south, the ring appears to be connected with a thin arm which is readily understood as the dust lane or off-axis shock created by the bar \citep{2012ApJ...747...60K, 2022ApJ...933..233F}. At the north, however, the ring appears to primarily connected to a thick arm, which lies further out than the northern counterpart of the off-axis shock. Although this arm still lies inside the corotation radius, its existence is not readily predicted in the canonical bar-driven inflow scenario. Regardless, the asymmetry in the two segments of the ring is likely seeded by this disparity in the outer gas steamers, which causes a different inflow rate and thus results in the lopsidedness and off-center geometry. 

The reason why the SW segment of the ring is predominantly ionized is unclear. One possibility is that the bar-induced shock is sufficiently strong to ionize the gas on the side where the gas density is relatively low. The higher temperature (and thus a higher sound speed) of the ionized gas would also bring the SW segment closer to the galactic center, as a higher thermal pressure of the gas can sustain stronger bar perturbation, as predicted by numerical simulations \citep{2012ApJ...747...60K}. 
Alternatively, perturbation by the M32 passage might have induced a gas inflow to the very center, temporarily enhancing the accretion rate and luminosity of the SMBH, subsequently resulting in the photoionization of the SW segment. In contrast, the NE arc might be more immune to this photoionization, due to either a higher density or a shield by the nuclear inflow itself. Interestingly, the magnetic field is found to be stronger in the SW segment \citep{2014A&A...571A..61G}, which may contribute to the higher ionization of the gas in this region. Moreover, the magnetic field also exhibits an asymmetric morphology and a lower inclination angle, providing further evidence of a previous interaction with M32. A better understanding of the ionization mechanism of the quasi-nuclear ring and nuclear spiral in M31 will be the subject of a future study.

The origin of the 10 kpc star-forming ring that is prominent in gas/dust is often discussed together with the quasi-nuclear ring. It is interesting to note that the outer Lindblad resonance (OLR) of the bar lies at 11.2 kpc, near the 10 kpc ring \citep{2006ApJ...650L..45B, 2006ApJ...638L..87G, 2009ApJ...699..486C}. Therefore, it is suggested that the outer 10 kpc ring could be a resonance ring \citep{2006MNRAS.370.1499A, 2015ApJ...805..183L}. However, the off-centered nature and the void within the ring are difficult to explain with a single bar-induced resonance. On the other hand, while the collisional ring scenario \citep{2006Natur.443..832B, 2006ApJ...638L..87G, 2014ApJ...788L..38D} provides a plausible explanation for these characteristics, it suffers from difficulties in terms of the high expanding velocity and uncertainties in the mass and orbit of M32 \citep{2015ApJ...805..183L}. Hence, a pre-existing resonance ring at the OLR, perturbed by the collision with M32, might offer a plausible explanation for the 10-kpc ring. In both our $N$-body and grid-based hydrodynamical simulations featuring a bar, a prominent 10 kpc ring emerges within several Myr after the collision. This demonstrates that the combination of a collision and a bar potential can reinforce the outer ring while efficiently clearing gas inside. Alternatively, a major merger with a mass ratio of 1: 4 \citep{2018MNRAS.475.2754H} could also yield the 10 kpc ring. However, a comprehensive, quantitative understanding of the origin of all the observed features, including the 10-kpc ring in M31 is beyond the scope of the present work and needs more constraints from multi-wavelength observations.

\section{Summary}
\label{sec: sum}
In this study, we have scrutinized the nature of the so-called nuclear ring (quasi-nuclear ring) in M31, originally defined by its dust emission but remaining poorly explored in its multi-phase gas components. We have utilized new JCMT CO(3-2) and IRAM CO(1-0) observations and archival data including VLA and FAST HI observations and CFHT/SITELLE spectroscopic mapping of the ionized gas, to examine the morphology and kinematics of the quasi-nuclear ring in unprecedented details. Several interesting signatures and properties of the quasi-nuclear ring are thus revealed:

\begin{itemize}
    \item  CO(3-2) and CO(1-0) emissions are unambiguously detected in the northeastern portion (NE arc) of the nuclear ring but are faint or absent in the southwestern portion (SW segment). Similarly, atomic hydrogen was only clearly detected along the NE arc, after careful subtraction of the foreground/background HI emission. 
    This asymmetry in the azimuthal distribution of the neutral gas is consistent with that seen in dust emission, but to a stronger degree, in the sense that the gas mass surface density exhibits a factor of $\gtrsim 5$ variation as the position angle of the ring varies. On the other hand, ionized gas is prominent in the SW segment. In addition, the total mass of the ring, estimated to be a few $10^6\rm~M_\odot$, is dominated by the molecular phase, which makes it similar to the case of the CMZ in our own Galaxy.

    \item The velocity field of the entire ring is derived for the first time by combining the neutral and ionized gas phases. 
    The line-of-sight velocities of the two phases are roughly consistent with each other over regions where they are both firmly detected. The azimuthal distribution of the line-of-sight velocity appears much flatter than that of a hypothetical rotating ring inclined at the same angle as the out disk of M31 ($77^\circ$) but is more compatible with the same ring viewed at an inclination angle of $\sim45^\circ$, at least for its NE arc. The SW segment, which is primarily manifested in ionized gas, exhibits a stronger scatter in the line-of-sight velocity as a function of the position angle.

\end{itemize}

The lopsided mass distribution of the ring strongly challenges the case of a genuine, coherently rotating ring.
Moreover, although a nuclear ring could be a natural product of the M31 bar, %\citep{2006MNRAS.370.1499A, 2007ApJ...658L..91B, 2018MNRAS.481.3210B}, 
we disfavor such a scenario because the off-centered morphology and the lower inclination angle inferred from the velocity field cannot be caused solely by a bar. 
To address the nature of the nuclear structures, we have performed custom hydrodynamical simulations of the M31--M32 collision, with and without a pre-existing bar, in an attempt to reproduce the observations. It is found that the %the kpc-scale ring-like structure at the center of M31 might have formed as a result of the recent interaction with M32. This interaction as a 
gravitational torque by M32 is most likely responsible for a tilt in the nuclear region, where asymmetric and eccentric spiral arms are generated under the restorative effect of the bar. At a certain moment and viewing angle, the projected tilted spirals can resemble the observed ring-like structure. %Therefore, this nuclear ring is transient and expected to deform gradually as M31 rotates and the system relaxes.
Therefore, the nuclear gas structure in M31 is most likely bar-induced asymmetric spirals with a significant tilt, rather than a persistent rotating ring. However, the imbalanced ionization state and off-centered appearance of this structure will be addressed in future work through detailed photoionization modeling combined with more elaborate hydrodynamical simulations.

\begin{acknowledgments}
This work was supported by the National Key Research and Development Program of China  (NO.2022YFF0503402 and No. 2022YFA1605000), MOST 2022YFA1605300, and the National Natural Science Foundation of China (grant 12225302). We thank Zixuan Feng for insightful discussions on collisional simulations with M32. Z.N.L. acknowledges support from the China National Postdoctoral Program for Innovation Talents (grant BX20220301) and the East Asian Core Observatories Association Fellowship. D.L. is a New Cornerstone investigator.

\end{acknowledgments}


\vspace{5mm}
\facilities{JCMT, IRAM 30m, VLA, FAST, CFHT/SITELLE %HST(STIS), Swift(XRT and UVOT), AAVSO, CTIO:1.3m, CTIO:1.5m,CXO
}


\software{astropy \citep{2013A&A...558A..33A,2018AJ....156..123A},  
          %Cloudy \citep{2017RMxAA..53..385F}
          %Source Extractor \citep{1996A&AS..117..393B}
          CASA \citep{2007ASPC..376..127M},
          GADGET-4 \citep{2021MNRAS.506.2871S},
          GILDAS \citep{2005sf2a.conf..721P},
          ORBS \citep{2012SPIE.8451E..3KM, 2015ASPC..495..327M}, 
          {\sc pluto} \citep{2007ApJS..170..228M},
          STARLIGHT \citep{2005MNRAS.358..363C},
          STARLINK \citep{2014ASPC..485..391C},
          yt \citep{2011ApJS..192....9T}
          }



\appendix

\section{Multiple Gaussian fitting of the HI data}\label{sec:HI_spec}

\setcounter{figure}{0}
\renewcommand\thefigure{A\arabic{figure}}

\begin{figure*}
	\centering
        \includegraphics[width=0.35\textwidth]{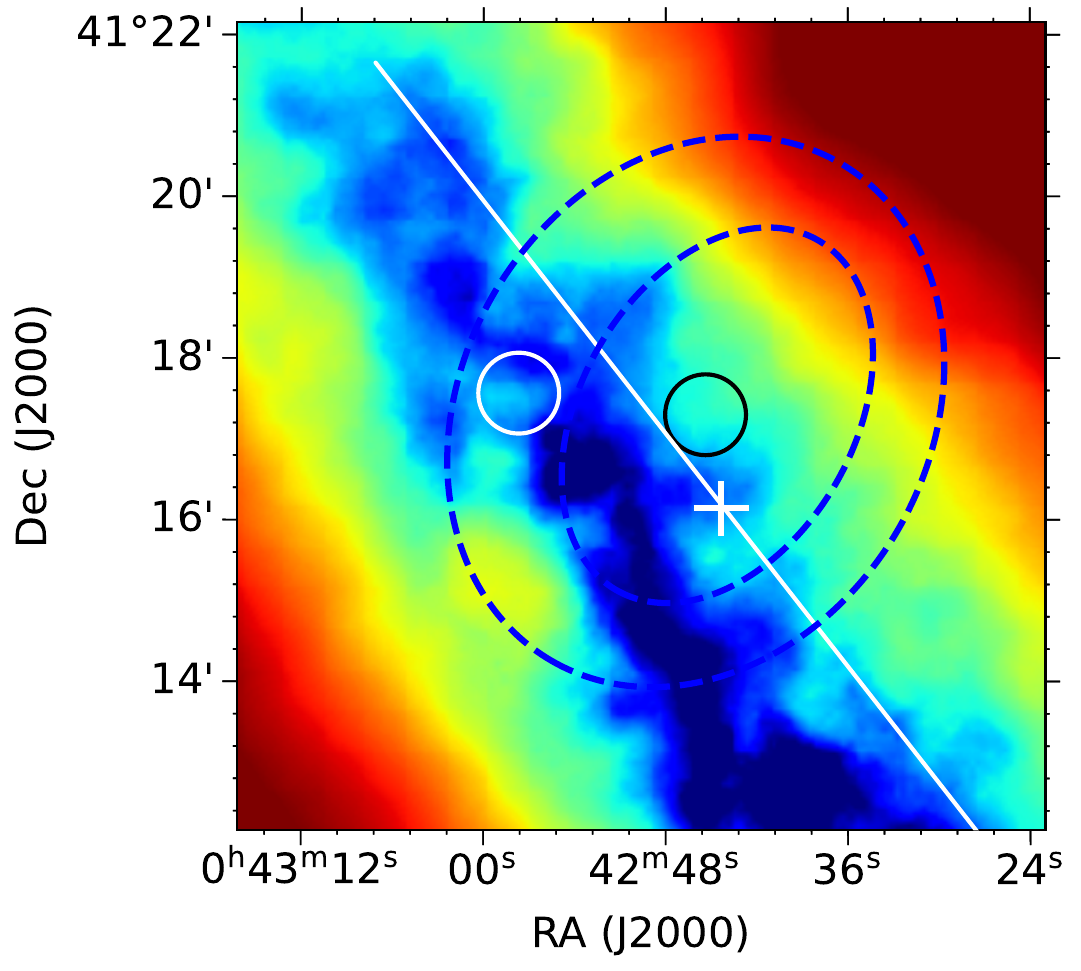}
        \includegraphics[width=0.64\textwidth]{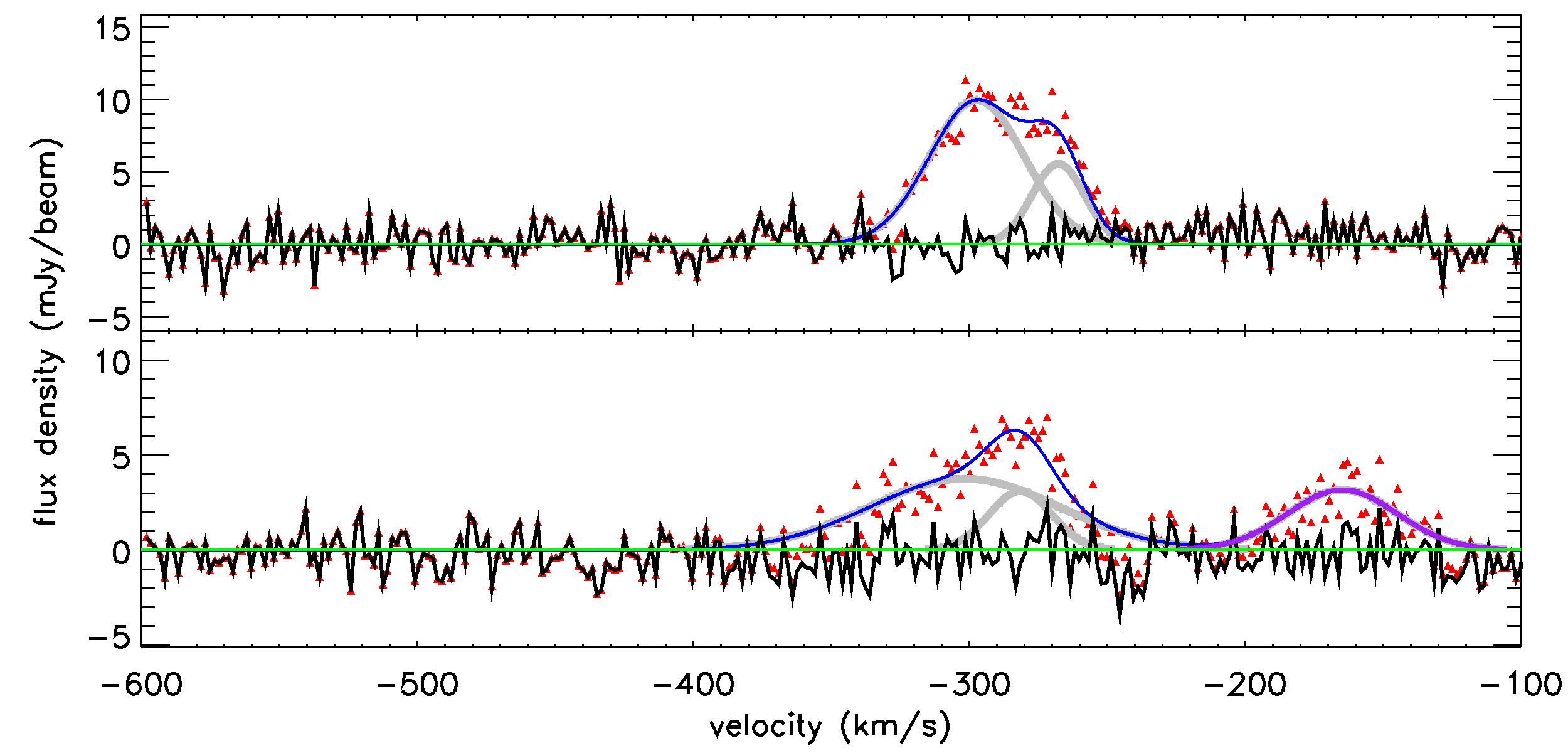}
	\caption{$Left$: The integrated intensity map of the HI data over the whole velocity range. The spectra within the two circles are extracted and displayed on the right panel. The white line indicates the path along the major axis, which was used to extract the PV diagram shown in Fig \ref{fig:PVD}. $Right$: Typical spectra on the NE arc of the quasi-nuclear ring (bottom panel, extracted from the white circle) and outside the ring region (top panel, extracted from the black circle). The red dots outline the spectrum, and the gray lines are the fitted multiple Gaussian components. The overall fitted spectrum is shown with a blue curve, and the component from the quasi-nuclear ring is highlighted with a bold purple curve. The residual is shown as a black line.
 \label{fig:HI}}
\end{figure*}

HI observations show multiple components in M31 \citep{2009ApJ...705.1395C} due to the diffuse nature of the atomic gas and the often limited spatial resolution. In particular, around $-300$ km s$^{-1}$, there is a broad ($\sim 50\rm~km~s^{-1}$ wide) velocity component that is ubiquitous in the entire nuclear region, as shown in Figure \ref{fig:HI} showing two typical spectra on and off the ring. %and in Figure \ref{fig:PVD} showing the PV diagram of the ring region. 
This component exhibits complex structures that plausibly arise from the foreground/background HI gas residing in the outer disk of M31. 

\begin{figure*}
	\centering
        \includegraphics[width=1\textwidth]{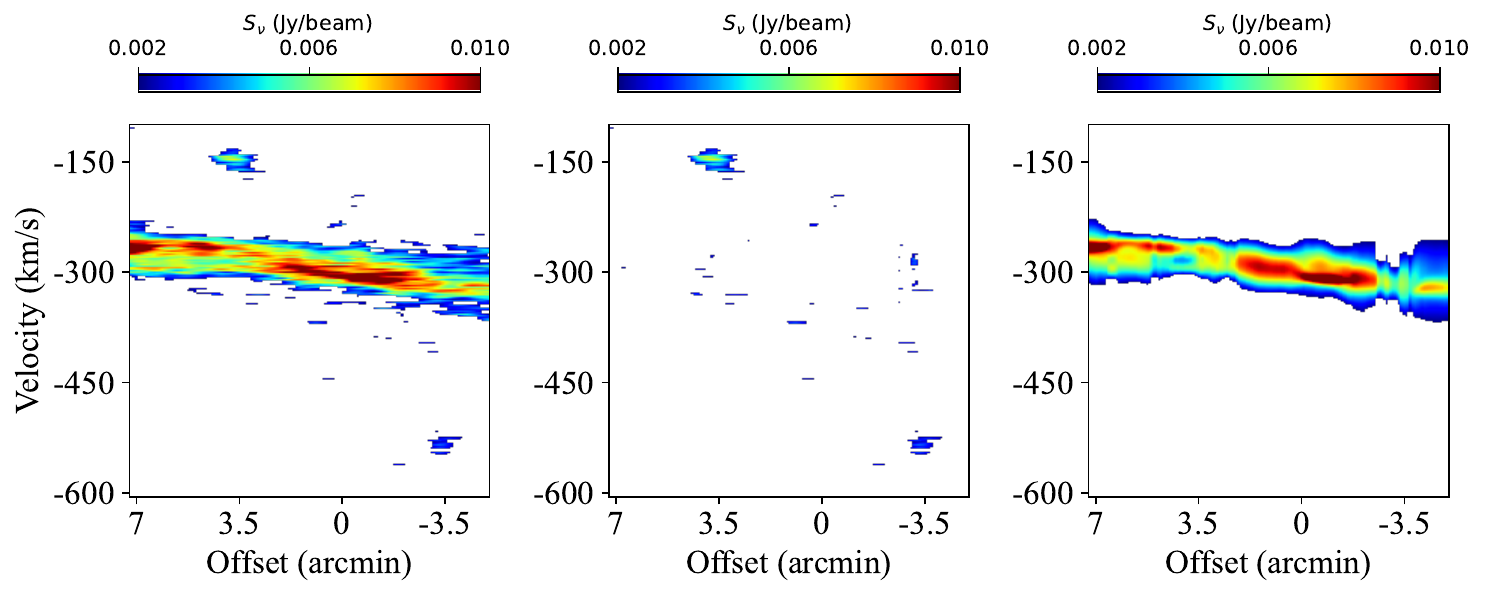}
	\caption{$Left$: The PV diagram of the HI data along the major axis shown in Figure \ref{fig:HI}, The offset is calculated from the M31 center. North is positive. The emission is dominated by a diffuse component at around $-300$ km s$^{-1}$. $Middle:$ The PV diagram of the HI cube after masking the dominant component around $-300$ km s$^{-1}$, which reveals the quasi-nuclear ring, in particular its NE arc around $-150$ km s$^{-1}$. $Right:$ The PV diagram of the residual cube, where the diffuse component around $-300$ km s$^{-1}$ is neatly separated.
 \label{fig:PVD}}
\end{figure*}

To separate the nuclear component, we use the Fit Multiple Gaussian components (FMG) tool based on the $\chi^2$ minimization procedure to fit multiple Gaussians to the combined HI data cube \citep{2023MNRAS.524.1169L}. This tool employs the Bayesian information criterion (BIC; \citealp{1978AnSta...6..461S}) to determine the adopted number of Gaussian components. 
Fitted models of the typical spectra on and off the quasi-nuclear ring are presented in Figure \ref{fig:HI}. The ring component at around $-150$ km s$^{-1}$ is distinct from the ubiquitous velocity component around $-300$ km s$^{-1}$. %Similarly, in the PV diagram along the nuclear ring (right panel of Figure \ref{fig:PVD}), the NE arc becomes prominent after removing the diffuse component. It is interesting to note that the ring is incomplete in the HI PV diagram. There is an apparent discontinuity in the SW segment of the ring, where the ionized gas is the brightest (Figures~\ref{fig:flux} and \ref{fig:optical}). %The nuclear ring component is distinct in the diagram, in addition to a component around -300 km s$^{-1}$ that is ubiquitous in this region. 
After this multiple Gaussian fitting, the ring component is neatly separated from the bulk diffuse component, as demonstrated in the PV diagrams shown in Figure \ref{fig:PVD}. 

\bibliography{M31_nuclear_ring}{}
\bibliographystyle{aasjournal}



\end{document}